\documentclass{emulateapj}
\usepackage{lscape}

\slugcomment{Version date: \today}

\shortauthors{Brittain, S. et al.}
\shorttitle{HD 100546}

\begin{document}
\title{Tracing the inner edge of the disk  around HD 100546 with Ro-vibrational CO Emission Lines}

\author{Sean D. Brittain}
\affil{Dept. of Physics \& Astronomy, 118 Kinard Laboratory, Clemson University, Clemson, SC 29634; sbritt@clemson.edu}
\author{Joan R. Najita}
\affil{National Optical Astronomy Observatory, 950 North Cherry Avenue, Tucson, AZ 85719; najita@noao.edu}
\author{John S. Carr}
\affil{Naval Research Laboratory, Code 7213, Washington, DC 20375; carr@nrl.navy.mil}

\begin{abstract}
  In this paper we present high resolution $4.7 \micron$ spectra of the isolated Herbig Be star HD~100546. HD~100546  has been the subject of intense scrutiny because it is a young nearby star with a transitional disk.  We observe the $\Delta$v=1 ro-vibrational CO transitions in order to clarify the distribution of warm gas in the inner disk. Modeling of the CO spectrum indicates that the gas is vibrationally excited by collisions and UV fluorescence. The observed emission extends from 13 - 100~AU. The inner edge of the molecular gas emission is consistent with the inner edge of the optically thick dust disk indicating that the inner hole is not simply a hole in the dust opacity but is likely cleared of gas as well.
  The rotational temperature of the CO is $\sim$1000K -- much hotter than the $\sim$200K CO in the otherwise similar transitional disk surrounding HD~141569. The origin of this discrepancy is likely linked to the brighter PAH emission observed toward HD~100546. We use the excitation of the CO to constrain the geometry of the inner disk and comment on the evolutionary state of the system. 

\end{abstract}

\keywords{ circumstellar matter --- line: profiles --- planetary systems:  protoplanetary disks --- stars: individual (HD 100546) --- techniques: spectroscopic}

\section{Introduction}
\object{HD 100546} is a $\sim$10 Myr, 2.4~M$_{\sun}$, isolated Herbig Be star (HBe; B9Ve) that has been particularly well studied because of its proximity to our Solar System (d=103$^{+7}_{-6}$ pc; van den Ancker 1997). The large disk encircling HD~100546 was first imaged in the NIR with ADONIS on the ESO 3.6m in LaSilla (Pantin et al. 2000).  Subsequent coronagraphic images with HST using NICMOS (Augereau et al. 2001), STIS (Grady et al. 2001) and ACS (Ardilla et al. 2007) as well as MIR interferometry with Magellan I (Liu et al. 2003) and the VLT (Leinert et al. 2004) revealed a $\sim$500AU disk. 

The disk around HD 100546 has also been imaged at 1.3mm with the 15m SEST telescope. The results indicate that the dust mass is $5\times10^{-4} \rm M_\sun$ assuming that the dust was 50K and optically thin (Henning et al. 1998). Two recent studies find a deficit of emission from dust within 10 AU of the star, (i.e. it is a transitional disk). Bouwman et al. (2003) model the thermal emission from the dust and  Grady et al. (2005) model the scattered light from the dust. The prevailing opinion by these groups is that the inner hole in the dust distribution is sculpted by an embedded companion that restricts the flow of material across its orbit.  In order to fit the SED of the disk, Bouwman et al. (2003) find that they need a small amount of dust in the inner 10~AU and a large puffed-up outer disk. They found the temperature of the dust in the puffed-up inner rim was 200~K. They also needed a scale height of 3.5AU to fit the MIR part of the SED. This is much higher than expected for 200~K gas 10~AU from a 2.4 M$_\sun$ star.  Grady et al. (2005) acquired FUV long-slit spectra of HD~100546 with STIS on HST. The spatial profile of the emission lines extends over several hundred AU due to scattering off the surface of the disk. However, this scattering is suppressed in the inner 13 AU suggesting this region has been largely cleared.  Thus HD~100546 is surrounded by a transitional disk with a massive dust component that extends from 13-500AU. While the dust component of the disk has been thoroughly characterized,  the gas component of the disk have not been as well studied. 

Emission from the rotational CO lines has been searched for but not detected resulting in no significant limit on the gas mass of the disk (see Loup et al. 1990; Nyman et al. 1992), but there are several lines of evidence indicating that the disk retains a significant gaseous reservoir. Firstly, HD 100546 is a Group Ia HBe star indicating that the disk is flaring (Meeus et al. 2001). Such flaring requires gas to support the dust. Secondly, Acke \& van den Ancker (2006) present a model of the spatial and spectral profile of the 6300 \AA [\ion{O}{1}] emission line. The shape of the line profile suggests that the emitting gas is in a disk and extends from 0.8-100 AU.

Thirdly,  there are several lines of evidence indicating that HD~1005456 is still accreting. While the profile of H$\alpha$ is highly variable (Vieira et al. 1999), the mean profile is similar to the synthetic profile of H$\alpha$ presented by Muzerolle et al. (2004) for the case of a Herbig Ae star accreting at 10$^{-9}$ M$_\sun$ yr$^{-1}$ and inclined by 60$\degr$. Balmer emission lines with absorption components red-shifted by as much as $\sim$220~km~s$^{-1}$, indicative of accreting gas, have also been observed (Guimar{\~a}es et al. 2006). In addition to the Balmer lines,  Deleuil et al. (2004) observed highly ionized species such as \ion{O}{6} with maximum velocities of $\sim$600 km s$^{-1}$ -- consistent with gas accreting magnetospherically.  While the accretion rate has not been determined, the veiling of the photospheric Ly$\alpha$ absorption line is a factor of 10 smaller than  HD~163296 and the flux of the red Ly$\alpha$ emission component is a factor of a 140 smaller than that of HD~104237 (Grady et al. 2005). These measurements led Grady et al. (2005) to conclude that the stellar accretion rate must be $\lesssim 10^{-9} \rm M_\sun  yr^{-1}$.  Unfortunately, the stellar accretion rate of Herbig Ae/Be stars is exceedingly difficult to measure (Muzerolle et al. 2004).  So while the accretion rate of HD~100546 has not been quantified, it is clear that this transitional disk retains a signficant reservoir of gas and is still accreting.   

In order to clarify the distribution of warm gas in the inner disk, and thereby explore the existence of the purported embedded companion, we study the CO ro-vibrational emission from the Herbig Be star HD~100546 which is found to have a strong fluorescent component. Like van der Plas et al. (2009), we find extended CO emission with a large inner hole.  The fundamental ro-vibrational CO emission lines have proven to be a reliable tool for probing the distribution of warm gas in the inner disk around young stars (e.g. reviews by Najita et al. 2000; 2007a and references therein). In particular, it has been shown that CO is a robust tracer of warm gas in the inner disk around Herbig Ae/Be stars with spectral types as early as B5 (Brittain et al. 2007; Blake \& Boogert 2004). In our earlier study of another Herbig Ae star with a purported inner hole, HD~141569, we showed that stellar UV irradiation of circumstellar CO can light up small amounts of gas, producing fluorescent emission even if the gas is cold (Brittain et al. 2007).  In that study, we were therefore able to show from the \emph{absence} of fluorescence emission at high velocity that there is a lack of CO within 10 AU of the star.  We employ a similar technique here to study the radial distribution of molecular gas in the HD~100546 disk. 
 
\section{Observations and data reduction}
The observations were obtained at the Gemini Observatory using the Phoenix echelle spectrograph (Hinkle et al. 1998; 2000; 2003) mounted on Gemini South (Table 1). The spectra were acquired with a 0.34$\arcsec$ slit that provided a resolving power of $\lambda/\Delta\lambda=50,000$ (6 km s$^{-1}$).  The spectral grasp of Phoenix is 1550 km s$^{-1}$ or roughly 10 cm$^{-1}$ at 5$\micron$. Thus 2-3 transitions per vibrational band fit within each setting. In order to sample a broad range of ro-vibrational energy levels we selected five settings centered at 2146 cm$^{-1}$, 2109 cm$^{-1}$, 2088 cm$^{-1}$, 2032 cm$^{-1}$ and 2011 cm$^{-1}$ (Fig. 1). The first four spectra were acquired January 14, 2006. The spectrum centered at 2011 cm$^{-1}$ was acquired on April 5, 2006. The seeing during these observations varied from 0.4$\arcsec$ - 0.6$\arcsec$ as measured from the spatial profile of the continuum of our spectra (Table 1). The signal to noise of the continuum of each spectrum is presented in Table 1 as well.

The slit was oriented in the default E-W direction. To cancel the sky background to first order the data were taken with an A-B-B-A nod sequence where A and B refer to E-W nod positions separated by the default 5$\arcsec$. A series of flats and darks were used to remove systematic effects at each grating setting. The 2-dimensional frames were cleaned of systematically hot and dead pixels as well as cosmic ray hits. Since the spectrum is curved along the detector the spectra must be rectified. A Gaussian curve is fit to the stellar point spread function (PSF) in each column and a second order polynomial is fit to the center of each fit. The columns are then shifted by the amount indicated by the polynomial. The standard deviation of the centroid of the rectified spectrum ranges from 6-15 mas (Table 1). A 1-d spectrum is then extracted from the rectified spectrogram. The width of the extraction window was 31 pixels (2.6$\arcsec$).

Telluric H$_2$O, CO$_2$, O$_3$ and CO have strong transitions in the regions where we observe (see, for example, the M-band spectra of HD~141569 presented in Brittain et al. 2007). This telluric absorption was corrected by observation of a bright star standard ($\alpha$ Cru) at a similar airmass. The wavelength regions with less than 50\% transmittance were not plotted to facilitate the presentation of the data. While accounting for the absorption of most molecules is reasonably straightforward, the depth of water lines can vary rapidly even from exposure to exposure. This can lead to mis-canceled telluric lines as well as discrepancies between the depth of water features in the telluric standard and science target. We discuss the specific lines affected by telluric corruption in \S3.1 and note them in table 2. 

The fluxes of the transitions were calculated assuming the star is M=3.8 (Malfait et al. 1998a). A spectrum of HD~100546 acquired with ISO-SWS reveals that the spectrum is flat over the range we are observing and 5.0$\pm$0.2 Jy (Malfait et al. 1998b). The ground based photometry was acquired between 1989 and 1992 while the ISO observation was taken in 1996. The fluxes inferred from both are consistent indicating that the flux at 5 $\micron$ is not variable at the few percent level.

\section{Analysis}
\subsection{Spectral Profile of the Emission lines}
A cursory glance at the high-resolution M-band spectrum of HD 100546 reveals a rich forest of CO emission lines (Fig. 1).  We detect $^{12}$CO lines originating from a wide range of upper vibrational levels (v$^{\prime}$=1 through v$^{\prime}$=8) as well as $^{13}$CO lines from the upper vibrational levels v$^{\prime}$=1 and v$^{\prime}$=2 (Tab. 2). The broad feature at 2148.9 cm$^{-1}$ is the Pf$\beta$ feature. We also see the wing of the Hu$\epsilon$ line begin at 2142 cm$^{-1}$. To extract the flux of CO lines superimposed on these features, we estimated the continuum by fitting to the regions on either side of the telluric line. There is a band of telluric O$_3$ running from about 2108-2112 cm$^{-1}$ which makes the spectrum particularly noisy. Additionally, there are telluric water features scattered throughout this region which can be highly variable. Lines whose telluric correction may be problematic are noted in table 2.  There is also an unidentified feature at 2030.9 cm$^{-1}$. It does not appear to be due to water or a mis-canceled telluric line.The shape of the line is similar to the CO lines suggesting that it arises from the same environment.

 A variety of line shapes are represented in the spectra due to line blending. Comparison of unblended emission lines demonstrates that they are spectrally resolved and have similar profiles (Fig. 2). This is consistent with the line shape presented by van der Plas et al. (2009) using higher-resolution spectroscopy (R$\sim$94,000). The flat topped shape of the lines is the tell-tale signature of an inclined rotating gas ring (e.g. Smak 1981). The velocity profile of the lines allows us to determine the distribution of the CO emission in the disk. The similarity of the line profiles indicates that the lines are formed over the same region, thus for the purposes of our modeling in this paper we assume that the gas originates in an azimuthaly symmetric disk in Keplerian rotation.  

The half width at zero\footnote{Zero is defined as less than 1-$\sigma$ above the continuum.}  intensity (HWZI) of the CO line is related to the maximum projected orbital velocity of the gas at the inner radius of emission and is 12$\pm$2 km s$^{-1}$ (Fig. 3).  To determine the inner radius of the gas, we correct for the instrument profile and disk inclination. The HWZI of the instrumental profile is 7 km s$^{-1}$.  The inclination of the disk reported from coronographic and interferometric imagery spans 42$\degr$-51$\degr$ (Pantin et al. 2000; Augereau et al. 2001; Grady et al. 2001; Ardilla et al. 2007; Liu et al. 2003; Leinert et al. 2004), however, most of these measurements converge at $\sim$50$\degr$ which we adopt for this paper. Thus we find that the maximum disk rotational velocity of the CO is 13$\pm$3 km s$^{-1}$. For gas in a Keplerian orbit about a 2.4 M$_\sun$ star, this velocity corresponds to a disk radius of 13$\pm$6 AU. 

\subsection{Spatial Profile of the Emission lines}
In addition to resolving the lines spectrally, it is possible to extract spatial information about the gas by centroiding the PSF of each velocity bin of the spectrum. We follow a similar extraction process as Acke \& van den Ancker (2006; see also Pontoppidan et al. 2008).  The slit for our observation was in its default E-W position as we did not expect to resolve the spatial extent of the CO lines with the typical seeing conditions at Gemini South ($\sim 0.5 \arcsec$). Unfortunately this orientation of the slit is not optimal for the position angle of HD 100546 (127$\degr$$\pm$5$\degr$; Grady et al. 2001).  To centroid the PSF of the emission lines, we begin by subtracting the stellar continuum from the spectrum. The columns that do not contain emission line flux from the source or telluric features are averaged. Then the average stellar PSF is scaled by the normalized standard star spectrum for each column across the array. The continuum subtracted image spanning 2142 - 2143 cm$^{-1}$ is presented in figure 4. This part of the spectrum was particularly clear of telluric features and the CO lines were minimally blended. The emission is clearly extended relative to the PSF of the continuum (Fig. 5) and the E-W sides of the PSF are Doppler shifted relative to one another. 

We fit a Gaussian curve to each column of the v=2-1 R(6) line to determine the centroid of the velocity bin (Fig. 6) and plot the spatial offset of each velocity bin relative to the centroid of the continuum. We find the maximum offset of the gas occurs between $\pm$ 5-7 km s$^{-1}$ (Fig. 7). The gas at maximum velocity and zero velocity is not significantly shifted from the centroid of the stellar position. This is exactly what one expects from gas in a rotating disk. The shift of the line centroid is nearly symmetric unlike the shift presented by van der Plas et al. (2009). The origin of this disparity is not clear and is the subject of ongoing work. 
In figure 8, we present a schematic of the disk imaged in the slit. The bins subtend 1.5 km s$^{-1}$ in both the radial and azimuthal direction, and the maximum velocity of each bin is labeled. The disk is inclined by 50$\degr$ and the disk semi-major axis is 127$\degr$ (Grady et al. 2005). Our slit was oriented E-W, so the disk semi-major axis was rotated by 37$\degr$ relative to the slit. Our data indicate that the gas emission originates in a rotating disk that extends out to $\sim$100AU (Fig. 5). We also find that the CO line profile at the east end of the slit is blue-shifted while the west end is red-shifted indicating that the CO is orbiting the star in a counter-clockwise direction. Acke \& van den Ancker (2006) found the same rotational direction for the 6300 \AA [\ion{O}{1}] emission line. Unlike the case of the 6300 \AA [\ion{O}{1}] emission line, we do not find that the CO extends inward of the dust. Acke \& van den Ancker (2006) suggest that the 6300 \AA [\ion{O}{1}] line results from the dissociation of OH. As in the case of HD~141569, this suggests that the inner disk is curiously rich in OH and depleted in CO. 

\subsection{Excitation of Emission Lines}
One can gain some insight into how the gas is excited by simply noting that emission lines from the vibrational levels v$^{\prime}$=1 through v$^{\prime}$=8 are observed.  The temperature necessary for populating these high-vibrational lines (T$_{\rm vib} \gtrsim 5500 K$) is higher than the dissociation temperature of CO. This suggests that the gas is UV fluoresced, because this process can lead to gas far from LTE such that it is much hotter vibrationally than rotationally (Krotkov et al. 1980). 

To determine determine the rotational temperature of the CO, we plot its excitation diagram. The flux of an optically thin ro-vibrational CO emission line is given by, 

\begin{equation}
F_{ij} =\frac{hc \tilde{\nu}_{ij} A_{ij} N_i}{4\pi d^2},
\end{equation}

\noindent where $N_i$ is the number of molecules in the upper state $i$ given by 

 \begin{equation}
N_i=\frac{Ng_i}{Q}e^{-E_i/kT}.
\end{equation}

\noindent Thus, we plot $ln(F_{ij}/(\tilde{\nu}A_{ij}g_i))\  vs. -E_i/k$ so that the negative reciprocal of the slope is equal to the temperature of the gas (Fig. 9). In table 3, we present the rotational temperature of each vibrational band which is  $\sim$ 1000K. UV fluorescence has also been observed in the transitional HAe star HD 141569 (Brittain et al. 2007). In contrast to HD~100546, the rotational temperature of the CO was only $\sim$200K even though the gas is distributed over a similar range of radii (Brittain et al. 2002; 2003; 2007; Goto et al. 2007). If the CO is thermalized, then $\sim$4\% of the molecules are in v=1 at 1000K. Thus collisional excitation may also be important and contribute to the population of v=1 unlike the case of HD 141569. To clarify the temperature, excitation and distribution of the gas, we generate a synthetic spectrum of the CO. 

\section{Spectral Synthesis}
To model the CO emission spectrum we calculate the relative vibrational populations of CO as a function of distance from the star.  We follow the prescription described in Brittain et al. (2007) for calculating the population of the ground electronic state vibrational levels, however,  we have extended this model for the current application by including collisional excitation.   

\subsection{Fluorescence}
CO has several bound excited electronic states. Of particular interest to this work is the 4$^{th}$ positive system of CO (transitions between the ground electronic state, $X^1\Sigma^+ $, and first excited electronic state, $A^1\Pi$) whose band origin is at 1545\AA\ and whose transitions range from 1300-2700\AA\ (Herzberg 1950).  To calculate the electronic excitation of CO, we determine the luminosity of the star in this wavelength region from a spectrum acquired with the International Ultraviolet Explorer (Valenti et al. 2000). The flux density at 1550~\AA \ is $\rm 2.5\times10^{-7}~ergs~s^{-1}~cm^{-2}~\micron^{-1}$.   Since the disk is highly inclined and nearby, we assume that there is only diffuse dust along the line of sight and thus use a normal ISM extinction law (Cardelli et al. 1989) and adopt A$\rm _V$=0.28 magnitudes (van den Ancker et al. 1998) to de-redden the spectrum.  

We set the inner edge of the disk at the radius we infer from the average spectral line profile (13 AU; Fig. 3) and adopt a 3.5 AU scale-height for the inner rim based on the SED modeling of this disk (Bouwman et al. 2003). The distance from the star determines the color temperature of the fluorescence spectrum that is emitted at a given radius (Krotkov et al. 1980; Brittain et al. 2007).  At larger radii, as the radiation field of the star grows more dilute, the spontaneous decay of the ro-vibrational levels becomes more significant and ÒcoolsÓ the levels more effectively, causing the vibrational temperature to drop. Therefore the fit to the relative strength of the transitions from high vibrational levels provides an independent constraint on the emitting radius. Eventually the field grows too dilute to populate the vibrational levels enough to give rise to observable lines.

The disk is divided into annuli 1AU wide and the UV flux from the star illuminates the front surface of the disk and the top of each subsequent annulus. At an inclination of 50$\degr$, only one side of the inner rim is observed as the other half is obscured by the disk itself (Fig. 8).  We therefore include only the flux from the projected area of the exposed portion of the inner rim to the flux of the first annulus for the calculation of the line profile. The disk is physically truncated above one scale height, and we assume the annuli are azimuthally symmetric.  Within a given ring we step along the line of sight into the inclined disk atmosphere in increments of  $\Delta\tau_0 = 0.2$ where $\tau_0$ is the line center optical depth of the strongest transition in the ground vibrational state.  

There are three free parameters we adjust to fit our model to the spectra. The turbulent velocity of the gas, the temperature at the inner edge of the disk and the radial dependence of the temperature. The turbulent velocity of the gas affects the brightness of the CO emission lines by determining the depth to which the pumping photons can penetrate. The temperature determines the relative brightness of the lines within each vibrational band. Finally the radial dependence of the temperature affects the line profile. The steeper the temperature profile, the larger the disparity between the shape of the low-J and high-J CO spectral line profiles.

The rms linewidth, $b$, is given by $(v_{thermal}^2  + v_{turbulent}^2)^{1/2}$ where $v_{thermal}$ is the component due to thermal broadening and $v_{turbulent}$ is the component due to turbulent broadening. The turbulent component of the intrinsic line width is a free parameter bounded by the requirement that the turbulent velocity not exceed the sound speed. Increasing $ v_{turbulent}$ has the effect of increasing the column of gas that is fluoresced and thus the flux of the CO. It also softens the double-peaked shape expected for gas in a rotating disk. We achieve our best fit to the CO line fluxes at a value of $v_{turbulent}$ = 2.0 km  s$^{-1}$. The UV flux from the star is attenuated in each step within the ring, the contribution to the emergent UV fluorescence emission is calculated, and the residual UV flux is allowed to excite the next step. For each ring we calculate the fluorescence of a total column density of 8 $\times$ 10$^{14}$ cm$^{-2}$. This radial column density along the line of sight to the star is selected so that the weakest transitions that we consider reach $\tau$=1. 

To calculate the angle of incidence of the UV entering the disk, we calculate the scale height of the disk assuming that it is at the temperature reflected by the rotational population of the CO. This angle ranges from 5$\degr$ at 13AU to 9$\degr$ at 100AU. Thus the perpendicular column density of the fluoresced CO is only $\sim$10$^{14}$ cm$^{-2}$. 

In our calculation we ignore the contribution by dust to the opacity of the gas. To check the reasonableness of this assumption, we consider the limiting case in which the CO/H ratio is $\sim$10$^{-6}$ and the gas/dust ratio is 100. To estimate the extinction, we treat the dust as spherical Mie scatterers (i.e. the cross section of grains larger than the extincted light is equal to their geometric cross section and the cross section of the smaller grains is $\pi a^3/\lambda$) with a grain size distribution $N(a) \propto a^{-2}$ and $0.01\micron\leq a \leq 10\micron \-$ the grain size distribution adopted by Bouwman et al. (2003) for their model of the SED of HD~100546. Based on these assumptions, we find that the total extinction at 1550 \AA over the UV fluoresced column along the line of sight to the star is 0.25 magnitudes. Thus it is reasonable to exclude the effect of dust in our calculation of the fluorescence of $^{12}$CO. Increasing the dust to CO ratio has the effect of reducing the luminosity of the CO emission.

The temperature of the gas is constrained by the rotational population of the vibrational bands. Since we initially ignore collisional excitation of the vibrational levels in this iteration of the model, the temperature does not impact the vibrational populations.  With this in mind, we assume the CO gas is in hydrostatic equilibrium, vertically isothermal and extends from 13 to 100 AU around a 2.4M$_\sun$ star (Table 2). The temperature profile was adjusted to fit the spectral profile of  the lines originating from $v^{\prime} \geq 2$ near 2142.7 cm$^{-1}$ (v=2-1 R(6), v=3-2 R(14) and v=4-3 R(23)) and 2112.5 cm$^{-1}$ (v=2-1 P(1), v=3-2 R(5) and  v=4-3 R(13); Fig. 10).  If the temperature profile is too steep, then the model did not fit the core of the high-J lines. The fiducial temperature affected the relative population of the rotational levels and was adjusted to maximize the fit of the high-J and low-J lines from a single level. We found that a temperature described by $T(R)=T_0\frac{R}{R_{in}}^{-0.3}$, where $T_0=1400 K$ and $R_{in}=13 AU$, provided a reasonable fit to our spectra. A steeper temperature profile resulted in a noticeable difference between the low and high J lines. A flatter temperature profile left the low-J lines under fit.   

To illustrate the effect of our key model parameters, $b$, $R_{in}$, and $T(0)$ we plot several fits to a subset of our data centered at 2143.2 cm$^{-1}$ and 2112.7 cm$^{-1}$ (Fig 11). We vary $b$ from 1-3 km s$^{-1}$,  $R_{in}$ from 6-19AU and $T(0)$ from 1000-1800 K. We also include a fit set at 200K for comparison with the rotational temperature of CO observed toward HD~141569. This is also the temperature of the dust at the inner edge of the disk around HD~100546. 

Given the vibrational population of the gas as a function of radius, we compute the synthetic spectrum. The intensity of the lines is given by $I=N_{vJ}hc\tilde{\nu}A_{vJ}$ because the lines are optically thin.  We assume that the intrinsic line profiles are Gaussian and turbulently broadened. Our model parameters are presented in table 3. The inner radius of the emitting region is determined from the HWZI of the emission lines. The outer radius of the disk is set large enough so that additional disk material no longer contributes to the luminosity of the CO ro-vibrational lines.  We convolve the spectrum of each annulus with a rotationally broadened line profile assuming the gas is in Keplerian orbit.  The stellar mass, disk inclination, and distance to the star have been culled from the literature  ($c.f. \S 1$) and are presented in table 3.  Next we sum the annuli and convolve the final spectrum with the instrument profile (a Gaussian profile with FWHM=6 km s$^{-1}$, the resolution of Phoenix). The seeing ranged from 0.4$\arcsec$-0.6$\arcsec$ and the slit was only 0.34$\arcsec$, thus it is necessary to account for slit losses. The disk is inclined by 50$\degr$, so only annuli beyond 30-50AU are affected. Gas from the region of the disk falling outside of our slit was excluded.  The flux of the spectrum is calculated by adopting a distance of 103 pc, and the model and flux calibrated data are presented in figure 10.  While the $^{12}$CO lines originating from v$\geq$2 and the $^{13}$CO lines are fit reasonably well, the model underestimates the flux of the v=1-0 lines. This indicates that collisional excitation is contributing to the population of v=1. This is not surprising for gas whose temperature is $\sim$1000K.

\subsection{Collisions}
To account for both collisional and fluorescent excitation, we have added collisional terms to our rate equation. We assume that the CO arises from a vertically homogenous, isothermal slab whose density has a radial dependence described by a power law (Table 2). In this slab, we assume that the hydrogen is primarily atomic so that the collisional rates are given by $k_{H-CO} = 7.57 \times 10^{-15}T(R)/1-e^{-3084/T} cm^3 s^{-1}  $ (see Najita et al. 1996 for a discussion of collisional excitation of CO).  Since UV fluorescence accounts for most of the line flux and these lines are optically thin, the collisionally excited gas must also be thin if the lines originate over the same surface area. Therefore we ignore radiative effects such as line trapping.  The density and temperature set the collisional rate.  The temperature of the gas was constrained by the relative strength of the rotational levels as described in $\S 4.2$.  The vibrational temperature of the UV fluoresced CO is set by the color temperature of the UV radiation field and the distance of the gas from the star. The v=2-1 lines are well fit with UV fluorescence alone, so we adopt a gas density that is low enough that v=2 is not appreciably populated thermally. A density profile of  n$\rm_H$=10$^{11}$ (R/13AU)$^{-1}$ cm$^{-3}$) provides a good fit. Much higher densities produce unacceptably strong v=2-1 emission. The surface density of the collisionally excited CO is 2.3$\times$10$^{15}$ cm$^{-2}$ or about an order of magnitude larger than the surface density of the fluoresced gas. There is a degeneracy between the thickness of the collisionally excited slab of CO and the density of the gas. The lower bound on the density is set by the critical density of rotational levels and is $\sim$10$^7$ cm$^{-3}$ (Kamp \& van Zadelhoff 2001).  As the density falls off, the efficiency with which v=1 is populated decreases and a thicker slab of CO is necessary to account for the strength of the v=1-0 lines. The resultant spectrum adopting a density at 13 AU of 10$^{11}$ cm$^{-3}$ is presented in figure 12 and the line fluxes are presented in table 2.

\subsection{$^{13}$CO}
In the ISM,  the ratio of $^{12}$CO to $^{13}$CO is $\sim$65 - 80 (e.g. Langer \& Penzias 1990), yet we detect relatively strong $^{13}$CO lines (\emph{cf.} the $^{13}$CO v=2-1 R(11) line at 2111.75 cm$^{-1}$ to the $^{12}$CO v=2-1 R6 line at 2142.5 cm$^{-1}$; figure 1). We find that the relative intensity of the v=1-0 and v=2-1 $^{13}$CO lines can be fit while ignoring collisions indicating that the $^{13}$CO is only excited by UV fluorescence (Fig. 10) and that the ratio of fluoresced $^{12}$CO to $^{13}$CO is roughly four. It is possible to drive the apparent abundance of $^{13}$CO up if the opacity of the  $^{12}$CO ro-vibrational lines is underestimated. However, the IR lines have oscillator strengths that are six orders of magnitude lower than the UV transitions, thus the observed IR lines must be very optically thin. In contrast, the electronic transitions are not optically thin. Indeed, if the CO were the only source of opacity in the disk, then the observed ratio of the isotologues should approach unity. The gas grows optically thick to the photons that excite the $^{12}$CO much sooner than to the photons that excite the $^{13}$CO. Thus the intensity of the $^{13}$CO emission continues to grow while the intensity of the $^{12}$CO remains relatively flat. This indicates that there must be an additional source of opacity in the disk. If we adopt the same disk parameters used to determine whether dust extinction is important for calculating the fluorescence of $^{12}$CO and also assume the ratio of $^{12}$CO to $^{13}$CO is $\sim$ 77 and the dust to gas ratio is constant over this column, then the total extinction along N($^{13}$CO)=8 $\times$ 10$^{14}$ cm$^{-2}$ is 19 magnitudes at 1550\AA\ (\emph{c.f.} \S 4.2) . This is sufficient to bring the ratio of $^{12}$CO to $^{13}$CO down to four. Unfortunately, the scaling factors used to arrive at this extinction are not particularly well constrained, so it is not possible to infer the proper gas/dust ratio in this part of the disk. Detailed modeling of the disk chemistry is necessary to constrain the composition of the gas and is beyond the scope of this paper. Whatever the case, it seems that dust extinction plays a role determining the relative intensity of the $^{13}$CO. This suggests that the surface region is not dust free. 

\subsection{The Emergent Spectrum}
Our spectral synthesis indicates that the gas extends from 13-100 AU, and 50\% of the flux emerges from beyond 33AU ( compare Figs. 5 and 13). This indicates that even small amounts of molecular gas ($\Sigma_{H}  = 1.7 \times 10^{-4}\  g \  cm^{-2}$ for CO/H=10$^{-6}$) can be detected from the outer disk of young stars if the disk is sufficiently flared.  The radial distribution of the CO emissivity that we infer from our spectral synthesis is consistent with the spatial distribution of the emission lines (Fig. 5). Roughly half of the gas originates within 0.2$\arcsec$ of the star. Since we are imaging a thin slice of the disk and our slit is not aligned along the semi-major axis (Fig. 8), this corresponds to 27AU from the star. The outer extent of the emission is 0.8$\arcsec$ which corresponds to 100AU.  Thus our fluorescence calculation, the velocity profile of the emission lines and the PSF of the CO emission are all consistent with CO ro-vibrational emission that extends from 13-100AU. The temperature of the gas that provides the best fit scales from 1400K at the inner edge and falls off as $T(r) \propto r^{-0.3}$.  This raises the interesting question of how gas more than 10 AU from a 10,000K star can be 1400K. More importantly, it can shed light on the unusual geometry of the disk around HD~100546.

\section{Discussion}
\subsection{Scale Height of the Inner Rim}
Modeling of the HD~100546 SED led Bouwman et al. (2003) to infer three dust distributions: a population of small grains ($M_{dust}=5.1\times10^{-9} M_{\sun}, a=0.1-10 \mu m$) distributed from 0.3-9.8 AU, a second population of small grains ($M_{dust}=6.7\times10^{-7} M_{\sun}, a=0.1-10 \mu m$) distributed from 9.8-43AU, and a population of large dust grains ($M_{dust}=6.5\times10^{-5} M_{\sun}, a=10-200 \mu m$) distributed from 28-380 AU. In this model, the density distribution of the dust is $\rho(r) \propto r^{-1}$ and the grain size distribution is $n(a) \propto a^{-2}$. The temperature of the dust at $\sim$10~AU required to fit the MIR luminosity is $\sim$200~K, but the scale height of hydrogen at this temperature is only $\sim$ 1~AU -- a third of the scale height of the inner rim inferred from the SED (Bouwman et al. 2003).

In comparison, the rotational temperature of the CO is $\sim$1400K at the inner edge of the outer disk (R=13~AU).  This is considerably hotter than the temperature of the dust at the same radius. It is expected that the temperature of the gas and dust should decouple in the low density region of the upper disk atmosphere and that the gas should be considerably hotter than the dust (Kamp \& van Zadelhoff 2001; Glassgold et al. 2004). The flux of the UV fluoresced CO suggests that the outer disk must be exposed to the UV field of the star.   If the gas here is in hydrostatic equilibrium and primarily atomic, then the scale height of the disk is $\sim$ 3AU at the inner edge -- similar to the scale height required by Bouwman et al. (2003) to reproduce the MIR portion of the SED.  

The temperature of the gas at the inner edge of the disk is much hotter than similarly situated gas in the nearly identical system HD~141569. In the case of HD~141569, the 200~K UV fluoresced CO in the disk extends outward from $\sim$9 AU (Brittain et al. 2007; Goto et al. 2007). The spectral types of the stars are similar, so if the temperature of the gas is determined by stellar irradiation (rather than accretion), the two gas temperatures should be similar. 

HD~100546 is a Group Ia source (Meeus et al. 2001) and HD~141569 is a Group II source (Acke \& van den Ancker 2004) suggesting that the disk around HD~100546 is more flared than the disk around HD~141569. This is consistent with the fact that the gas is warmer in the HD~100546 disk. The increased flaring of the disk around HD~100546 exposes the disk to the star more directly leading to enhanced heating, but why should the disk be warmer in the first place? The difference could be due to a greater abundance of small grains and/or PAHs in the HD~100546 disk.  Calculations by Kamp et al. (2005) indicate that PAHs and small grains play an important role in heating disk atmospheres around T Tauri stars through enhanced photoelectric heating of the gas. Ongoing calculations also indicate that PAHs drive up gas temperatures in disks around HAeBes (I. Kamp, private communication). Indeed, the luminosity of the PAH emission from HD~100546 is 20 times greater than the luminosity of the PAH emission from HD~141569 (Acke \& van den Ancker 2004). Further, the PAH emission from HD~100546 is distributed from 12$\pm$3 AU to beyond 50 AU (Habart et al. 2006; Geers et al. 2007), so it is radially coincident with the CO emission. The high temperature of the CO is qualitatively consistent with the temperature of the top most molecular layer in the disk model of Kamp et al. (2005).
 
\subsection{The Inner Hole}
Modeling of the SED indicates a small amount of dust in the inner 10 AU of the disk (Bouwman et al. 2003). Is the presence of such (potentially extincting) dust consistent with the observation  of UV fluoresced gas in the outer disk (R$\gtrsim$ 10~AU)? In order to fit the NIR portion of the SED, Bouwman et al. (2003) invoke the presence of 10$^{25}$ g of dust in the inner disk with a scale height of H/R$\sim$0.085. Since the size of the small grains scales as $a^{-2}$ rather than $a^{-3.5}$ and the maximum size of the 'small dust' is 10$\micron$ rather than $\sim$ 0.25 $\micron$ as in the ISM, the extinction in the UV is greatly diminished. However, the extinction can still be significant at 0.15$\micron$ and thus shadow the inner wall of the disk from the pumping photons.  To check this, we calculate the extinction of the UV flux from the star again assuming the dust can be treated as Mie scatterers. Under this assumption, the UV extinction from the star to the inner rim along the disk midplane is $\sim$0.8 magnitudes. Thus the dust necessary to produce the NIR excess of the SED will not significantly attenuate the UV flux needed to fluoresce the CO in the outer disk (13 AU $\leq$ R $\leq$ 100 AU).  

Next we use the absence of broad wings on the emission lines to place a tight constraint on the amount of gas in the inner disk. In figure 14 we have presented a synthetic line profile of the gas distributed from 0.3 AU$\leq$R$\leq$13 AU. We assume that the temperature and density of the gas follow the same power-law described in \S4.1 and \S4.2. The entrance angle of the UV flux from the star into the disk ranges from $\sim$1.3$\degr$ at 0.3AU to $\sim$5$\degr$ at 13AU. We then scale the vertical surface density of the CO to set the upper limit on the flux. The 3$\sigma$ upper-limit on the vertical surface density of CO ranges from 5$\times$10$^{13}$ cm$^{-2}$ at 0.3 AU to 2$\times$10$^{14}$ cm$^{-2}$ at 13AU. The average vertical surface density is 1.3$\times$10$^{14}$ cm$^{-2}$. 

Adopting this upper limit and CO/H=10$^{-6}$ (e.g. Thi \& Bik 2005) indicates that the surface density of gas in the inner 13AU is $\lesssim$2$\times$10$^{-4}$ g cm$^{-2}$. Thus the mass of gas in the inner disk is $\lesssim$3 $\times$10$^{25}$ g.  Bouwman et al. (2003) find that 10$^{25}$ g of dust in the inner disk is necessary to explain the NIR excess, so our results indicate that the gas-to-dust ratio in the inner disk $\sim$3 -- not nearly enough gas needed  for a gas-rich disk like the minimum solar mass nebula. While our measurements only constrain the CO column and not the total gas column, it seems unlikely that we  are simply seeing evidence of the destruction of CO in the inner disk as CO has been observed into the dust destruction radius around stars as early as B5 such as HD~259431 (Brittain et al. 2007). 

The low upper limit we find on the column density of CO in the inner  
disk contrasts with the model for HD~100546 presented in Acke et al.  
(2005).  There, they assumed that the [OI] 6300\AA\ emission they observe  
is produced by the photodissociation of OH in the very surface layers  
of a high column density disk.  In their model for HD~100546, the  
vertical mass column density at 1AU is 10$^4$ g cm$^{-2}$.  For a typical  
midplane abundance of CO ($\sim$10$^{-4}$ relative to hydrogen), the  
corresponding vertical column density of CO is many orders of  
magnitude larger than our upper limit.  Since CO is chemically robust  
and able to self-shield at CO column densities larger than $\sim$10$^{15}$ cm$^{-2}$  
(e.g., Thi \& Bik 2005), our upper limit on the CO column within 13  
AU suggests that the gas mass within this radius is much less than  
assumed in the model of Acke et al. (2005).  The [OI] emission  
observed may be produced from the photodissociation of OH, but the  
total gas reservoir in this region appears likely to be much more  
tenuous than they have assumed.

In their study of the CO emission in HD100546, van der Plas et al.  
(2009) found a similar result to that obtained here, that the CO emission  
is restricted to radii $>$11AU.  They interpret this result as  
indicating that the disk within 11 AU is gas rich, but depleted in CO,  
perhaps as a result of photodissociation and other chemical pathways.   
Since photodissociation will only affect the disk surface layers (as a  
result of CO self-shielding), an (as yet unspecified) chemical pathway  
would be needed to remove the CO in the rest of the disk.  It would be  
interesting to see if future models of inner disk chemistry can  
provide support for such a scenario.
  
\subsection{An Embedded Companion?}
The identification of planets in the process of forming will enable \emph{in situ} studies of planet formation that will advance our understanding of this process in the same way that the study of stellar accretion revolutionized our understanding of star formation. Unfortunately, direct imaging of a forming gas-giant planet remains out of reach with current technology, thus it is necessary to look for indirect signposts of this process.

An important tool for tracing the evolution of circumstellar disks is the measurement of the SED of young stars (e.g. Strom et al. 1989; Hillenbrand et al. 1992; Malfait et al. 1998). The differences in the SEDs of young stars have led some to propose that they reflect an evolutionary sequence from optically thick disks to transitional disks (i.e. disks with optically thick outer disks and optically thin inner disks) to optically thin disks on a timescale commensurate with planet formation (e.g. Malfait et al. 1998). The most common interpretation of the distinctive transitional disk morphology is that it reflects dynamical sculpting by an embedded planet (e.g. Skrutskie et al. 1990; Marsh \& Mahoney 1992; Calvet et al. 2002; Rice et al. 2003;  Bouwman et al. 2003; Quillen et al. 2004; D'Alessio et al. 2005; Calvet et al. 2005; Furlan et al. 2006). 

As exciting as it is to use SEDs of young stars to identify embedded planets, it is crucial to keep in mind that other physical processes can give rise to transitional disks (see for example Najita et al. 2007b for a summary of such processes).  Observation of spectrally resolved warm molecular emission is an important tool for understanding the relative distribution of gas and dust in the inner disk, and thus breaking the degenerate interpretations of transitional SEDs. 

There are at least four scenarios that can give rise to a transitional disk. One possibility (scenario A) is the preferential agglomeration of dust in the inner disk (Strom et al. 1989; Fig. 15A). Such a process can lead to a decrease in the continuum opacity while retaining its original gas surface density. If this scenario occurs, then the gas will fill the optically thin region of the transitional disk and show up in bright emission lines.  A second possibility (scenario B) is the photoevaporation of the disk by the star (Clarke et al. 2001; Fig. 15B). As the disk dissipates, the accretion rate falls to the photoevaporation rate at some point in the disk (the critical radius). The material interior to this radius is isolated from material exterior to this point and the inner disk is emptied. Consequently, the optically thin region of the disk is completely devoid of dust and gas and the outer disk will have a very low surface density (generally $\lesssim$ 0.1 g cm$^{2}$ at 10AU). Transitional disks are also expected to form in response to the formation of gas giant planets (Skrutskie et al. 1990; Fig. 15C; scenario C). The formation of a Jovian-mass planet will dynamically reduce the flow of gas and dust from the outer disk across its orbit into the inner disk (Lin et al. 1993; Lubow et al. 1999; Lubow 2006). The gap-crossing material enters into the inner disk along accretion streams but does not fill this region, thus the NIR excess is diminished and the star reveals a transitional SED. The gas in these streams will build up near the star as it accretes and will glow in a narrow ring. Thus the outer radius of the gas emission will be smaller than the inner radius of the optically thick portion of the disk, and hot UV fluoresced gas will be observed.  Finally, a related scenario (D) occurs when the gas giant planet grows large. A planet with a mass that exceeds  5M$_{\rm Jupiter}$ entirely stop accretion across its orbit (Lubow et al. 1999; Fig. 15D). When this occurs the material in the inner disk will quickly accrete onto the star. There will be no warm gas emission from the inner disk, but cool UV fluoresced gas will be instead observed from the massive outer disk. This scenario can be distinguished from case B by the surface density of the outer disk.

HD~100546 is an interesting system with an inner hole that is plausibly explained by an embedded companion.  Several lines of evidence indicate that the inner radius of the dust is $\sim$10~AU (Bouwman et al. 2003; Grady et al. 2005). The sharp drop in the surface density of CO at 13 AU indicates that the inner disk is not gas-rich and dust-poor (scenario A). The possibility of ongoing (but lower) accretion and the presence of OI in the inner disk supports scenario C. However, the lack of CO in the same region seems to imply that the material in the inner disk is tenuous suggesting a scenario closer to D. 

\section{Conclusion}
Our excitation+disk synthesis model does a reasonable job of reproducing the observed flux and line profile of the CO spectrum. There is an apparent asymmetry in the data that we do not account for and whose origin is unclear. Accounting for this asymmetry is subject of ongoing work.  Our modeling of the CO spectrum from this source indicates that the molecular gas is truncated at $\sim$13 AU.  The absence of molecular gas in the inner disk suggests that the material in the inner disk is tenuous. Such a hole is consistent with the presence of a relatively massive companion.  We also find that the temperature of the gas at the inner edge is $\sim$1400 K. If this gas is in hydrostatic equilibrium, then the scale height of the disk is $\sim$3 AU, consistent with the presence of a puffy inner rim inferred from SED fitting.  This provides additional evidence for a few theoretical expectations from recent disk modeling.

First, we find that the temperature of the gas and the dust are decoupled (see for example, Glassgold et al. 2004). Second, the large scale height required for the gas and dust at 13AU indicates that this hot gas and small dust grains are well mixed. We have also highlighted the possible importance of PAH heating on the temperature of the gas in the disk. While the stars HD 141569 and HD 100546 are similar and have disks with similar sized inner holes, the temperature of the gas is radically different (200 K vs 1400 K). The difference in the gas temperature between the two systems argues that it is not simply grain settling that is responsible for the difference in the SEDs of the two systems.  It is possible that the measured difference in the gas temperature is a consequence of a differing abundance of small grains and PAHs in the two systems.

Finally, our data illustrate the utility of UV fluoresced CO for studying the giant planet forming region of the disk (R \textgreater 10 AU). Indeed half of the flux we observed comes from beyond 30 AU. Thus the non-detection of CO emission around stars with a significant UV continuum can place a severe constraint on the existence of molecular gas in the disk. We also find that UV fluorescence effects can complicate attempts to use $^{13}$CO detections to infer the surface density of emitting CO. While the IR lines are optically thin, the UV transitions are extremely optically thick.

\acknowledgments
Based on observations obtained at the Gemini Observatory, which is operated by the Association of Universities for Research in Astronomy, Inc., under a cooperative agreement with the NSF on behalf of the Gemini partnership: the National Science Foundation (United States), the Particle Physics and Astronomy Research Council (United Kingdom), the National Research Council (Canada), CONICYT (Chile), the Australian Research Council (Australia), CNPq (Brazil) and CONICET (Argentina). The Phoenix infrared spectrograph was developed and is operated by the National Optical Astronomy Observatory.  The Phoenix spectra were obtained as part of programs GS-2005B-C-2 and GS-2006A-C-17. The authors wish to thank Inga Kamp for helpful discussions about the role of PAHs in circumstellar disks. S.D.B. acknowledges support for this work from the National Science Foundation under grant number AST-0708899 and NASA Origins of Solar Systems under grant number NNX08AH90G. Basic research in infrared astronomy at the Naval Research Laboratory is supported by 6.1 base funding.

{\it Facilities:} \facility{Gemini:South (Phoenix)}

\clearpage

\begin{deluxetable}{lllcccc}
\tablenum{1}
\tablewidth{0pt}
\tablecaption{Log of Observations}

\tablehead{ \colhead{Date} &\colhead{Telescope/Instrument} & \colhead{Spectral Grasp}  &  \colhead{Integration}  & \colhead{Seeing} & \colhead{S/N} & \colhead {Centroid Scatter}\\
\colhead{}  & \colhead{} & \colhead{cm$^{-1}$} & \colhead{minutes} & \colhead{arcseconds} & \colhead{}& \colhead{milliarcseconds}} 
\startdata
2006 Jan 14 & Gemini S./PHOENIX & 2141.0 - 2152.0 & 20  & 0.40 & 102 &  6.12\\
 & & 2104.5 - 2113.5 & 12 & 0.42 & 76 & 7.56 \\
 & & 2084.5 - 2093.0 &  20 & 0.54 & 41 & 9.78\\
 & & 2027.0 - 2038.0 & 12 & 0.59 & 43 & 14.9\\
2006 Apr 05 &  & 2016.0 - 2006.0 & 16 & 0.53  & 53 & 10.9 \\
\enddata
\end{deluxetable}

 \clearpage
\LongTables 
\begin{landscape}

\begin{deluxetable}{llllccl}
\tablenum{2}
\tablewidth{0pt}
\tablecaption{Line Fluxes}

\tablehead{\colhead{Species} & \colhead{Vibrational} &\colhead{Line} & \colhead{$\tilde{\nu}$} & \colhead{Flux} &\colhead{Flux} & \colhead{Notes} \\	
\colhead{} & \colhead{band} &\colhead{} & \colhead{} & \colhead{Data} &\colhead{Model} & \colhead{} \\	
\colhead{} & \colhead{} & \colhead{} & \colhead{cm$^{-1}$} & \colhead{10$^{-14}$ erg s$^{-1}$ cm$^{-2}$} & \colhead{10$^{-14}$ erg s$^{-1}$ cm$^{-2}$} & \colhead{}}

\startdata			
$^{12}$CO	&	v=1-0	&	P(31)	&	2008.53	&	13.9$\pm$0.2	&	14.02	&		\\
$^{12}$CO	&	v=1-0	&	P(30)	&	2013.35	&	 7.6$\pm$0.5	&	7.92		&		\\
$^{12}$CO	&	v=1-0	&	P(27)	&	2027.65	&	 9.4$\pm$0.5	&	11.67	& Positioned on the wing of a strong water line		\\
$^{12}$CO	&	v=1-0	&	P(26)	&	2032.35	&	10.3$\pm$0.3	&	10.96	&		\\
$^{12}$CO	&	v=1-0	&	P(25)	&	2037.03	&	14.0$\pm$0.4	&	14.33	&		\\
\\
\hline
\\
$^{12}$CO	&	v=2-1	&	P(26)	&	2006.78	&	3.1$\pm$0.2	&	3.08	&		\\
$^{12}$CO	&	v=2-1	&	P(25)	&	2011.42	&	3.1$\pm$0.2	&	3.17	&		\\
$^{12}$CO	&	v=2-1	&	P(21)	&	2029.66	&	3.9$\pm$0.3	&	4.33	&		\\
$^{12}$CO	&	v=2-1	&	P(8)		&	2085.34	&	4.7$\pm$0.7	&	4.97	&		\\
$^{12}$CO	&	v=2-1	&	P(7)		&	2089.39	&	4.6$\pm$0.3	&	5.23	&		\\
$^{12}$CO	&	v=2-1	&	P(2)		&	2109.14	&	2.3$\pm$0.3	&	1.42	& crowded telluric region		\\
$^{12}$CO	&	v=2-1	&	P(1)		&	2112.98	&	2.7$\pm$0.4	&	2.58	&		\\
$^{12}$CO	&	v=2-1	&	R(6)		&	2142.47	&	6.6$\pm$0.2	&	5.78	&		\\
$^{12}$CO	&	v=2-1	&	R(7)		&	2146.00   	&	8.6$\pm$0.2	&	9.80	&	blended with v=3-2 R(15) line	\\
$^{12}$CO	&	v=2-1	&	R(8)		&	2149.49	&	4.9$\pm$0.5	&	5.49	&	Pf$\beta$	\\
\\
\hline
\\
$^{12}$CO	&	v=3-2	&	P(15)	&	2030.16	&	3.9$\pm$0.3	&	3.93	&		\\
$^{12}$CO	&	v=3-2	&	P(14)	&	2034.41	&	4.0$\pm$0.4	&	4.31	&		\\
$^{12}$CO	&	v=3-2	&	P(1)		&	2086.60	&	1.4$\pm$0.3	&	0.56	& superimposed on a telluric water line		\\
$^{12}$CO	&	v=3-2	&	R(5)		&	2112.29	&	3.1$\pm$0.1	&	2.95	&		\\
$^{12}$CO	&	v=3-2	&	R(14)	&	2142.72	&	3.9$\pm$0.2	&	4.88	&		\\
$^{12}$CO	&	v=3-2	&	R(15)	&	2145.92	&	8.6$\pm$0.2	&	9.80	&	blended with v=2-1 R(7) line	\\
$^{12}$CO	&	v=3-2	&	R(16)	&	2149.08	&	3.0$\pm$1.0	&	4.80	&	Pf$\beta$	\\
\\
\hline
\\
$^{12}$CO	&	v=4-3	&	P(13)	&	2012.73	&	7.5$\pm$0.3	&	7.95	&	blend	\\
$^{12}$CO	&	v=4-3	&	P(8)		&	2033.14	&	2.4$\pm$0.2	&	2.52	&		\\
$^{12}$CO	&	v=4-3	&	R(6)		&	2089.22	&	2.6$\pm$0.3	&	2.44	&		\\
$^{12}$CO	&	v=4-3	&	R(7)		&	2092.68	&	3.9$\pm$0.1	&	3.68	&	blend	\\
$^{12}$CO	&	v=4-3	&	R(11)	&	2106.14	&	3.0$\pm$0.1	&	3.34	& 		\\
$^{12}$CO	&	v=4-3	&	R(12)	&	2109.41	&	1.6$\pm$0.3	&	3.43	& crowded telluric region		\\
$^{12}$CO	&	v=4-3	&	R(13)	&	2112.65	&	2.8$\pm$0.1	&	3.51	&		\\
$^{12}$CO	&	v=4-3	&	R(23)	&	2142.95	&	3.1$\pm$0.2	&	2.89	&		\\
$^{12}$CO	&	v=4-3	&	R(25)	&	2148.56	&	2.0$\pm$0.2	&	2.05	&	Pf$\beta$	\\
\\
\hline
\\
$^{12}$CO	&	v=5-4	&	P(8)		&	2007.14	&	1.6$\pm$0.2	&	1.90	&		\\
$^{12}$CO	&	v=5-4	&	P(7)		&	2011.09	&	2.1$\pm$0.2	&	1.64	&		\\
$^{12}$CO	&	v=5-4	&	P(6)		&	2015.00	&	3.0$\pm$0.2	&	2.21	&		\\
$^{12}$CO	&	v=5-4	&	P(2)		&	2030.31	&	0.7$\pm$0.2	&	0.56	&		\\
$^{12}$CO	&	v=5-4	&	R(13)	&	2085.88	&	6.0$\pm$0.8	&	5.58	&	blend	\\
$^{12}$CO	&	v=5-4	&	R(14)	&	2089.05	&	2.0$\pm$0.3	&	2.37	&		\\
$^{12}$CO	&	v=5-4	&	R(15)	&	2092.17	&	1.9$\pm$0.1	&	2.40	&		\\
$^{12}$CO	&	v=5-4	&	R(21)	&	2110.15	&	0.9$\pm$0.5	&	1.99	& crowded telluric region		\\
$^{12}$CO	&	v=5-4	&	R(33)	&	2141.94	&	1.0$\pm$0.2	&	0.71	&		\\
$^{12}$CO	&	v=5-4	&	R(34)	&	2144.33	&	0.8$\pm$0.2	&	0.51	&		\\
$^{12}$CO	&	v=5-4	&	R(35)	&	2146.69	&	0.6$\pm$0.1	&	0.49	&		\\
\\
\hline
\\
$^{12}$CO	&	v=6-5	&	R(5)		&	2032.83	&	2.1$\pm$0.4	&	2.16	&		\\
$^{12}$CO	&	v=6-5	&	R(6)		&	2036.25	&	1.4$\pm$0.3	&	1.14	&		\\
$^{12}$CO	&	v=6-5	&	R(23)	&	2088.79	&	0.9$\pm$0.3	&	1.40	&		\\
$^{12}$CO	&	v=6-5	&	R(24)	&	2091.54	&	1.0$\pm$0.3	&	1.03	&		\\
$^{12}$CO	&	v=6-5	&	R(31)	&	2109.71	&	0.4$\pm$0.1	&	0.54	&		\\
\\
\hline
\\
$^{12}$CO	&	v=7-6	&	R(5)		&	2006.48	&	0.7$\pm$0.2	&	0.71	&		\\
$^{12}$CO	&	v=7-6	&	R(6)		&	2009.87	&	0.6$\pm$0.2	&	0.74	&		\\
$^{12}$CO	&	v=7-6	&	R(13)	&	2032.56	&	1.1$\pm$0.2	&	1.03	&		\\
$^{12}$CO	&	v=7-6	&	R(14)	&	2035.66	&	1.2$\pm$0.4	&	1.10	&		\\
$^{12}$CO	&	v=7-6	&	R(32)	&	2084.86	&	2.4$\pm$0.3	&	1.55	&		\\
$^{12}$CO	&	v=7-6	&	R(44)	&	2110.55	&	0.7$\pm$0.1	&	0.89	& 		\\
\\
\hline
\\
$^{12}$CO	&	v=8-7	&	R(15)	&	2012.10	&	0.6$\pm$0.3	&	0.77	&		\\
$^{12}$CO	&	v=8-7	&	R(23)	&	2034.92	&	0.7$\pm$0.3	&	0.50	&		\\
\\
\hline
\\
$^{13}$CO	&	v=1-0	&	P(21)	&	2012.21	&	1.6$\pm$0.3	&	1.77	&		\\
$^{13}$CO	&	v=1-0	&	P(17)	&	2029.24	&	6.6$\pm$0.1	&	6.69	&		\\
$^{13}$CO	&	v=1-0	&	P(16)	&	2033.42	&	1.9$\pm$0.2	&	1.66	&		\\
$^{13}$CO	&	v=1-0	&	P(1)		&	2092.39	&	0.5$\pm$0.2	&	0.18	& 		\\
$^{13}$CO	&	v=1-0	&	R(2)		&	2106.90	&	0.3$\pm$0.2	&	0.65	&		\\
$^{13}$CO	&	v=1-0	&	R(3)		&	2110.44	&	0.9$\pm$0.1	&	0.88	&		\\
$^{13}$CO	&	v=1-0	&	R(13)	&	2144.03	&	2.0$\pm$0.2	&	2.00	&		\\
$^{13}$CO	&	v=1-0	&	R(15)	&	2150.34	&	2.0$\pm$0.2	&	2.15	&	superimposed on the Pf$\beta$ line	\\
\\
\hline
\\
$^{13}$CO	&	v=2-1	&	P(11)	&	2028.91	&	1.2$\pm$0.4	&	1.08	&		\\
$^{13}$CO	&	v=2-1	&	P(9)		&	2036.80	&	1.4$\pm$0.4	&	0.96	&		\\
$^{13}$CO	&	v=2-1	&	R(4)		&	2088.47	&	0.8$\pm$0.4	&	0.72	&		\\
$^{13}$CO	&	v=2-1	&	R(5)		&	2091.92	&	1.1$\pm$0.2	&	0.97	&		\\
$^{13}$CO	&	v=2-1	&	R(11)	&	2111.88	&	0.5$\pm$0.1	&	1.31	& on the edge of a deep telluric line		\\
$^{13}$CO	&	v=2-1	&	R(23)	&	2148.07	&	1.2$\pm$0.2	&	0.88	&	superimposed on the Pf$\beta$ line	\\
\enddata
\end{deluxetable}
\clearpage
\end{landscape}

\begin{deluxetable}{lll}
\tablenum{3}
\tablewidth{0pt}
\tablecaption{Rotational Temperature of CO}
\tablehead{ \colhead{Species} & \colhead{v} & \colhead{Temp. (K)} }\\

\startdata
$^{12}$CO	&	1	&		900$\pm$200			\\
$^{12}$CO	&	2	&		880$\pm$50			\\
$^{12}$CO	&	3	&		800$\pm$100			\\
$^{12}$CO	&	4	&		1100$\pm$100			\\
$^{12}$CO	&	5	&		960$\pm$50			\\
$^{12}$CO	&	6	&		800$\pm$100			\\
$^{12}$CO	&	7	&	\nodata \\
$^{12}$CO	&	8	&	\nodata \\
$^{13}$CO	&	1	&		860$\pm$180 \\
$^{13}$CO	&	2	&		1100$\pm$200 \\
\enddata

\end{deluxetable}

\begin{deluxetable}{llll}
\tablenum{4}
\tablewidth{0pt}
\tablecaption{Model Parameters}

\tablehead{ \colhead{Parameter} & \colhead{Value} &\colhead{Description} }\\
\startdata
M$_{\star}$ & 2.4 M$_\sun$ & Stellar mass  \\
i & 50$\degr$ & Inclination	 \\
d & 103 pc & Distance to star  \\
$\Delta$R & 6.0 km  s$^{-1}$ & Resolution of instrument \\

\\
\hline
\\

R$_{\rm in}$ & 13 AU& Inner edge of disk  \\
R$_{\rm out}$ & 100 AU & Outer edge of disk  \\
v$_{\rm turbulent}$ & 2.0 km s$^{-1}$ & Turbulent broadening  \\
T & 1400 K & Fiducial temperature of CO at 13~AU \\
$\alpha$ & 0.30 & Power law of rotational temperature  \\
n$_0$(H) & 10$^{11}$ cm$^{-3}$ & Fiducial density at 13~AU  \\
$\beta$ & 1.35 &	Power law of density  \\
X &  4	& $^{\rm 12}$CO/$^{\rm 13}$CO ratio  \\
Y & 34	& C$^{\rm 16}$O/C$^{\rm 18}$O ratio  \\

\enddata
\end{deluxetable}

\clearpage

\clearpage
\begin{figure} 
\epsscale{0.85}
\plotone{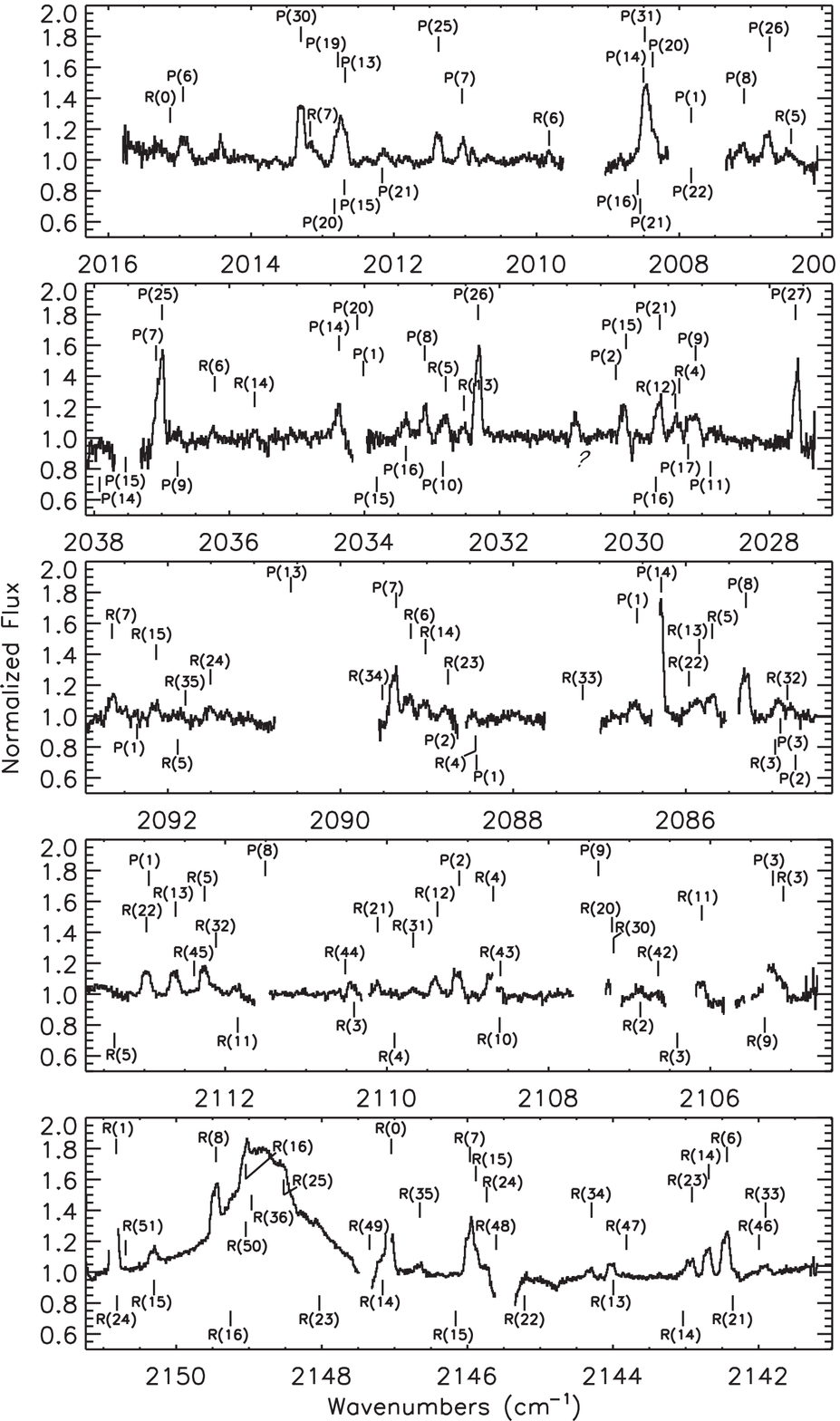}
\caption{M-band spectrum of HD 100546. The narrow emission lines are from CO. The broad emission feature at 2148.8 cm$^{-1}$ is the Pf $\beta$ line. The $^{12}$C$^{16}$O $\Delta$v=1 lines are labeled above the spectrum such that the v=1-0 lines are labeled at 1.95 units, the v=2-1 lines are labeled at 1.85 units, and so forth through v=7-6. The $^{13}$C$^{16}$O  $\Delta$v=1 lines are plotted below the spectrum where the $^{13}$C$^{16}$O v=1-0 lines are plotted at 0.8units, the $^{13}$C$^{16}$O v=2-1 lines are plotted at 0.7 units. We also label the position of the $^{12}$C$^{18}$O v=1-0 lines (the labels are at 0.6 units) which are not observed. The gaps in the spectrum are of regions where the atmospheric transmittance is less than 50\%. }
\end{figure}

\clearpage

\begin{figure} 
\epsscale{1.0}
\plotone{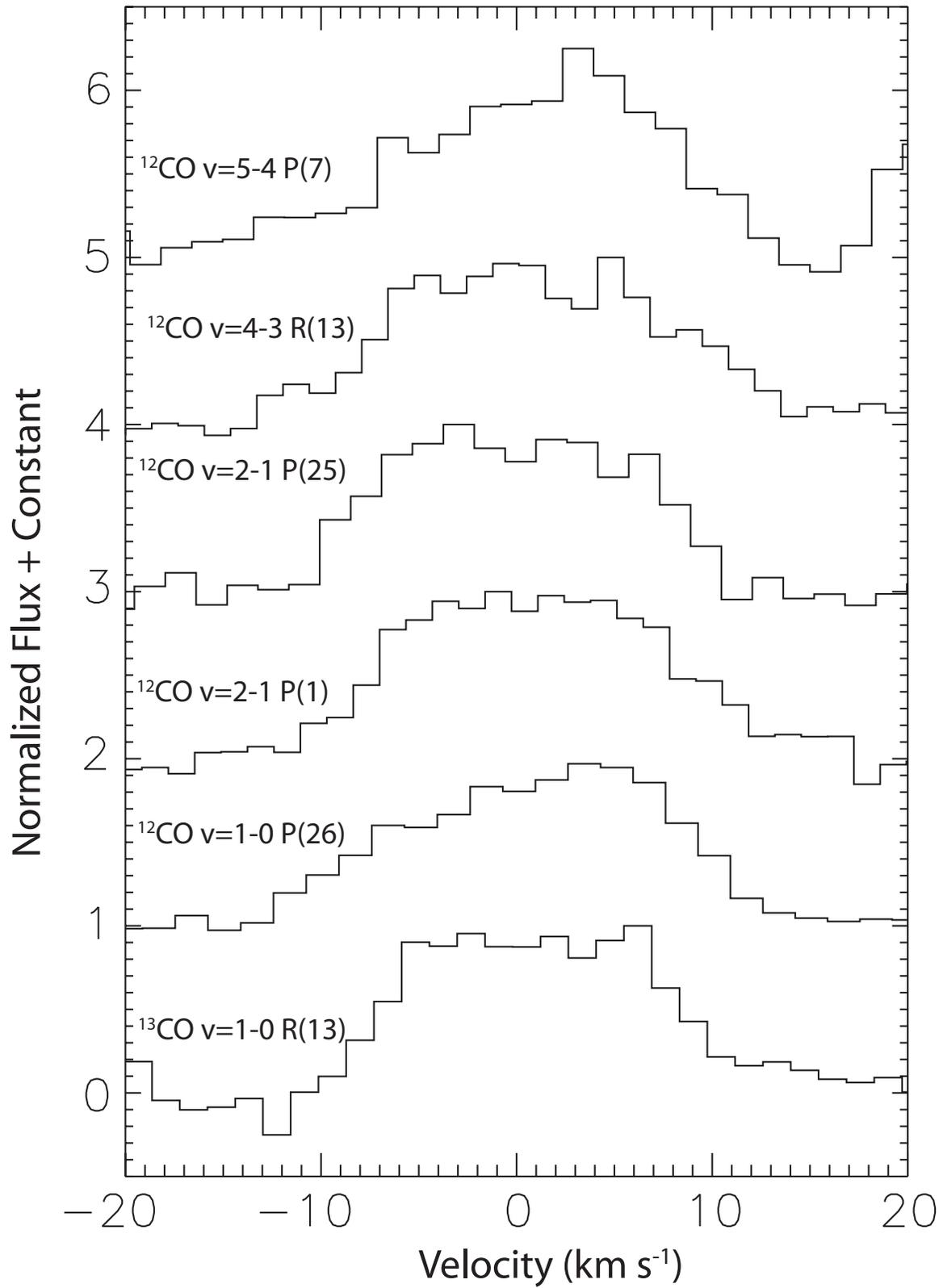}
\caption{Comparison of kinematic profile of isolated emission lines. These lines are a selection of isolated CO lines showing that the lines are resolved and flat topped -- indicative of gas in an inclined rotating disk (e.g. Smak 1981). The similarity of the line shapes indicates that they are formed over the same region of the disk.}
\end{figure}

\begin{figure} 
\epsscale{0.8}
\plotone{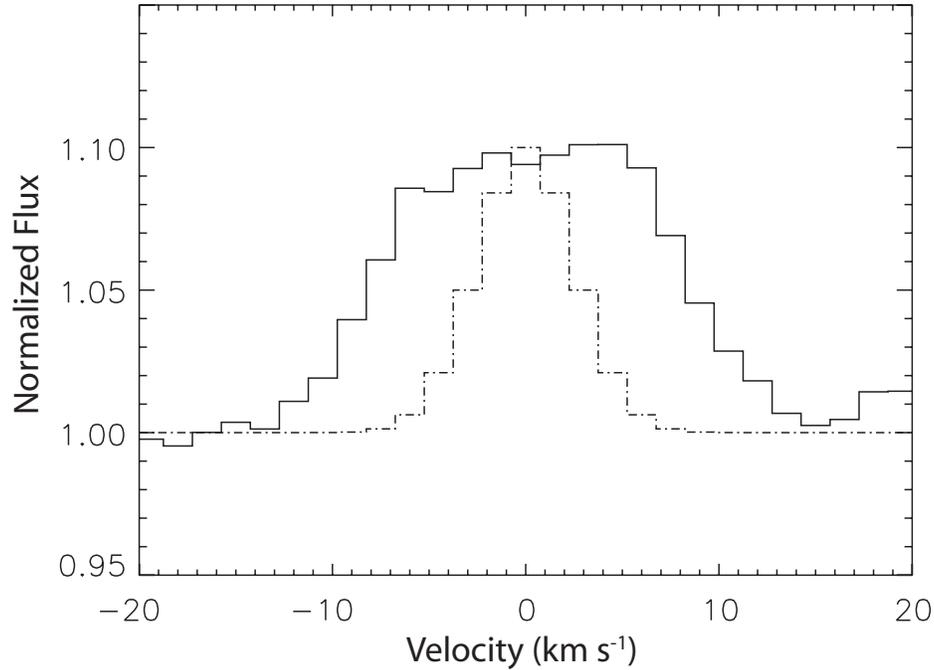}
\caption{Average profile of isolated CO emission lines. We have averaged six unblended emission lines presented in figure 2 (solid line) and overplotted a Gaussian curve with a FWHM=6 km s$^{-1}$ representing the instrument profile (dot-dashed line). The HWZI of the CO lines is 12 $\pm$ 2 km s$^{-1}$ and the HWZI of the instrument profile is 7 km s$^{-1}$.  The CO lines are resolved and at our S/N nearly symmetric.  There appears to be a slight excess on the red wing of the profile from 12-18 km s$^{-1}$, the origin of which is unclear.  The flat topped shape of the line is indicative of gas in  a rotating disk. The deprojected maximum velocity of the CO gas is 13 km s$^{-1}$ which corresponds to an inner radius of 13 AU. The peak separation corresponds to the outer extent of the gas.} 
\end{figure}

\begin{figure}
\epsscale{0.8}
\plotone{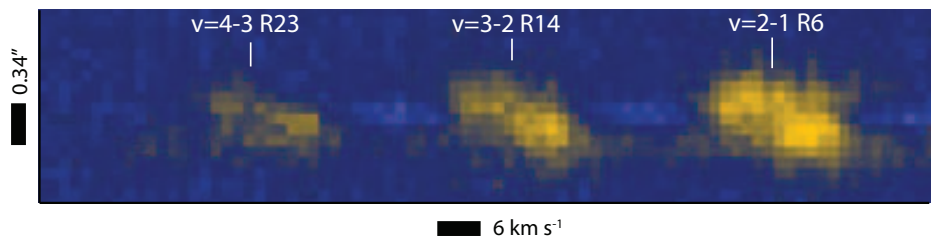}
\caption{Detail of three continuum subtracted emission lines. The average profile of the continuum was scaled by the transmittance of the telluric standard. This scaled and averaged PSF was subtracted from each column of the spectrogram revealing the spatially and spectrally resolved emission lines. Centered at 2142.6 cm$^{-1}$, this spectrogram highlights the spatial and spectral structure of these lines. The spatial dimension of the plot is in the vertical direction (0.085 $\arcsec$ per pixel). The east end of the plot is at the top. The spectral dimension is in the horizontal direction (1.5 km s$^{-1}$ per pixel). The black bar by the y-axis corresponds to the spatial resolution and the black bar below the x-axis corresponds to the spectral resolution of the spectrum. The red side of the line profiles (to the right) appears to be brighter than the blue side.}
\end{figure}

\begin{figure} 
\epsscale{0.9}
\plotone{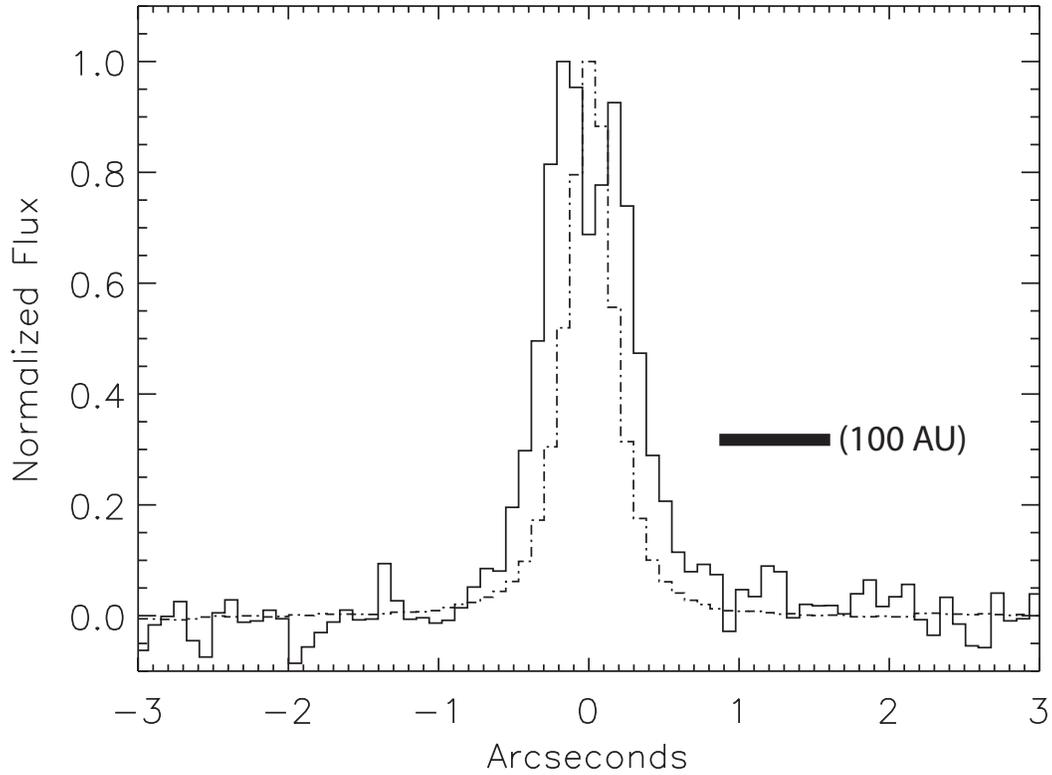}
\caption{Point Spread Function of the CO emission. The angular extent of the normalized v=2-1 R6 CO line (solid profile) is compared to the point-spread-function of the source continuum (dot-dashed profile). The CO emission is clearly spatially extended relative to the continuum. The plate scale of Phoenix on Gemini South is 0.085$\arcsec$ per pixel, so each pixel corresponds to 8.8AU at the distance to HD~100546. However, the slit was misaligned with the semi-major axis of the disk by 37$\degr$, so 8.8AU on the plane of the sky corresponds to a distance of 11AU along the disk. The inset, horizontal bar represents a distance of 100AU along the disk. }
\end{figure}
\clearpage

\begin{figure}
\epsscale{0.5}
\plotone{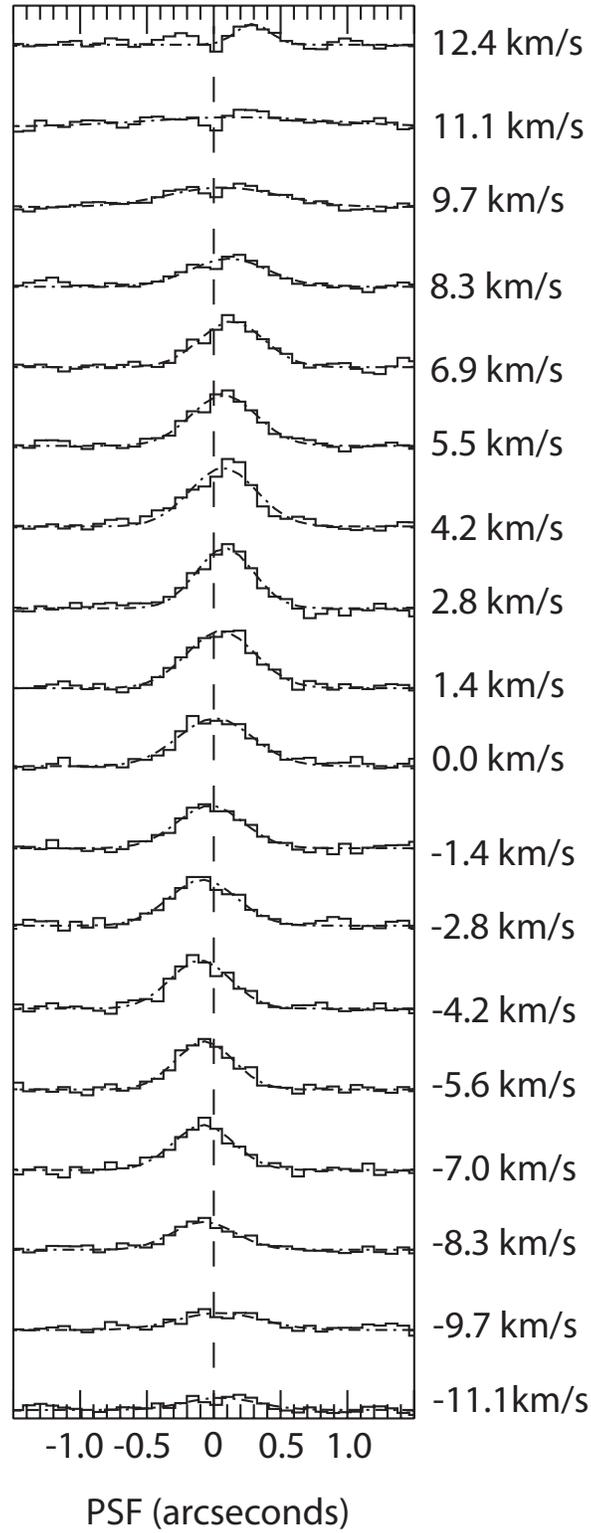}
\caption{Spatial cut of the v=2-1 R(6) line. Each column of the normalized, continuum subtracted v=2-1 R(6) line is plotted. The wavenumber scale has been converted to velocity space so that the center of the line is at 0 km s$^{-1}$. The negative shift is to the west (lower on the chip). The center of the stellar PSF is labeled with a dashed line and each profile is fit with a Gaussian curve.}
\end{figure}

\clearpage

\begin{figure}
\epsscale{0.8}
\plotone{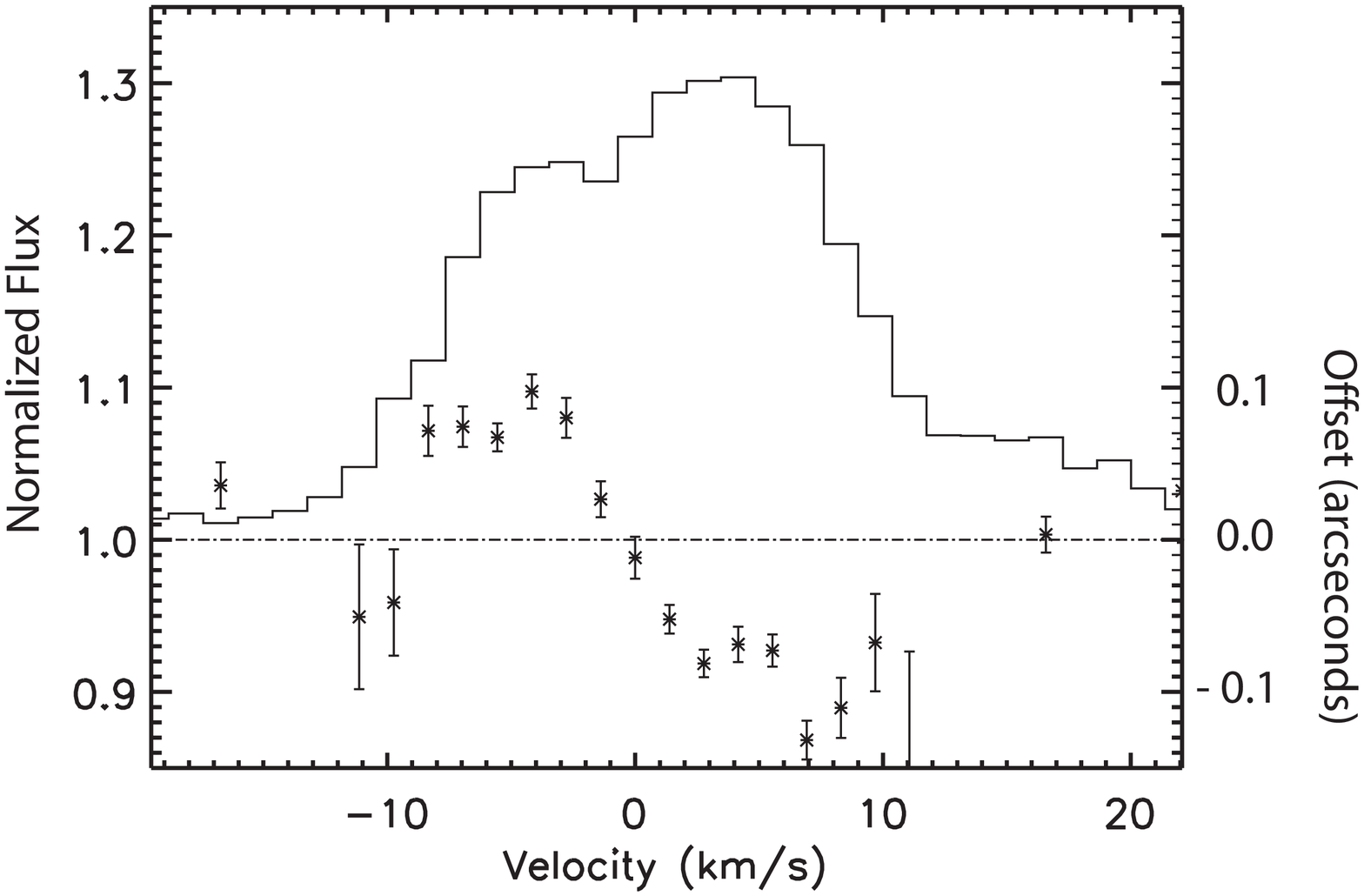}
\caption{Centroid plot of v=2-1 R(6) line at 2146 cm$^{-1}$. The normalized line is plotted in velocity space. The spatial offset of the center of the Gaussian curve fit to each column is also plotted. The negative offset corresponds to the west. There is significant scatter because the profiles of the lines are not symmetric Gaussian curves. Some of this deviation may be due to the error in the subtraction of the stellar continuum. The maximum offset occurs at $\sim$ 6 - 7 km s$^{-1}$ and is roughly 0.1$\arcsec$. }
\end{figure}

\begin{figure}
\epsscale{0.5}
\plotone{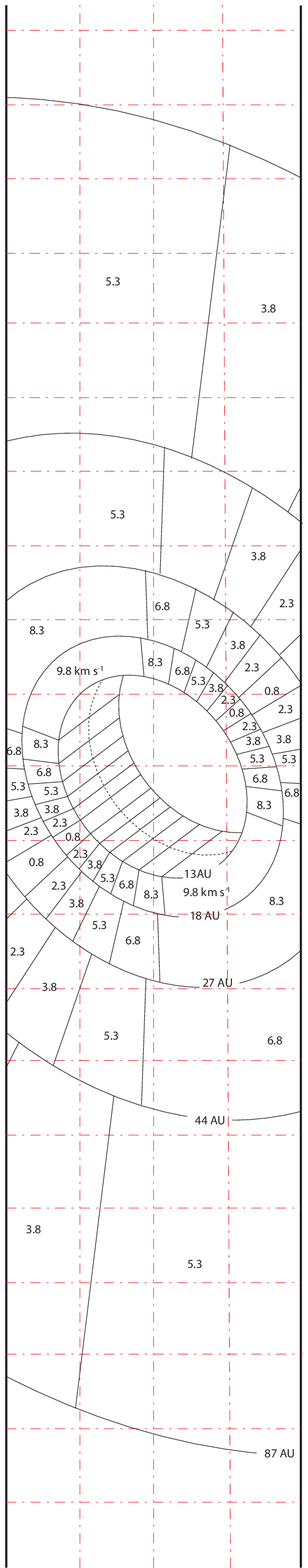}
\caption{Schematic of the disk. The image of the disk on the slit is plotted by adopting an inclination of 50$\degr$, position angle of 127$\degr$ and an E-W slit orientation. The pixels of the instrument are represented by the red dot-dashed lines. The segments of the disk are 1.5 km s$^{-1}$ line of sight projected velocity bins (in both the radial and azimuthal directions) where the label of each bin is the maximum velocity of that bin. The inner radius of each annuli is also labeled. The inner R=87AU of the disk covers 9 pixels in this orientation, consistent with the PSF of the CO lines and the prediction from our fluorescence model.}
\end{figure}

\clearpage

\begin{figure} 
\epsscale{0.8}
\plotone{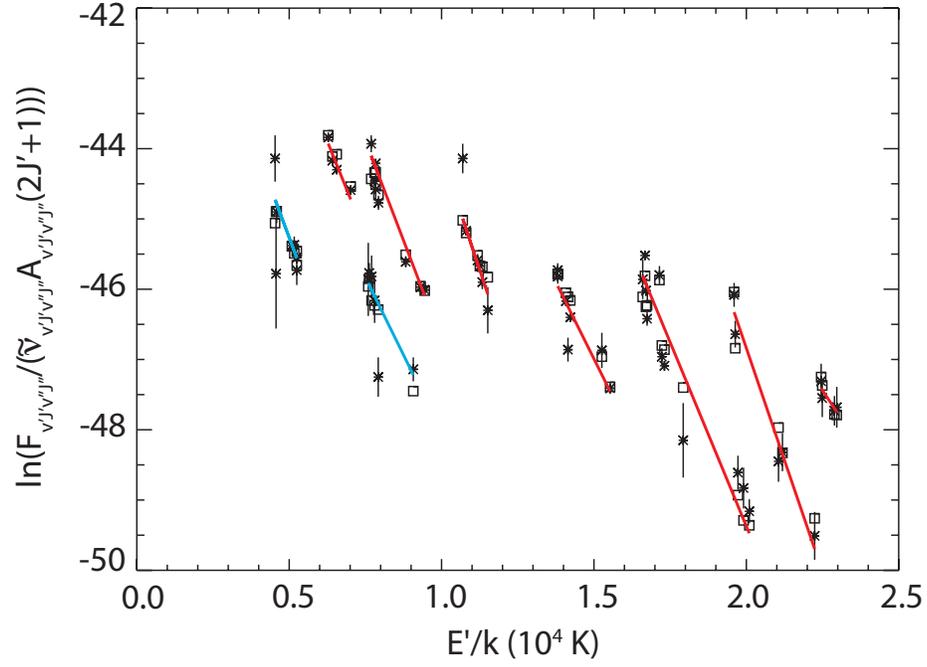}
\caption{Excitation diagram of CO emission lines. $ln(F/(\tilde{\nu}A(2J^{\prime}+1)))$ is plotted versus $E^{\prime}/k$ so that the rotational temperature of the gas is equal to negative reciprocal of the slope of the weighted least squares fit.  Values extracted from the data are plotted with asterisks while values extracted from our model are plotted with squares. The fit to the $^{13}$CO lines is cyan and the fit to the  $^{12}$CO lines is red. The temperatures of the fits are presented in table 3.}
\end{figure}

\clearpage

\begin{figure} 
\epsscale{1.0}
\plotone{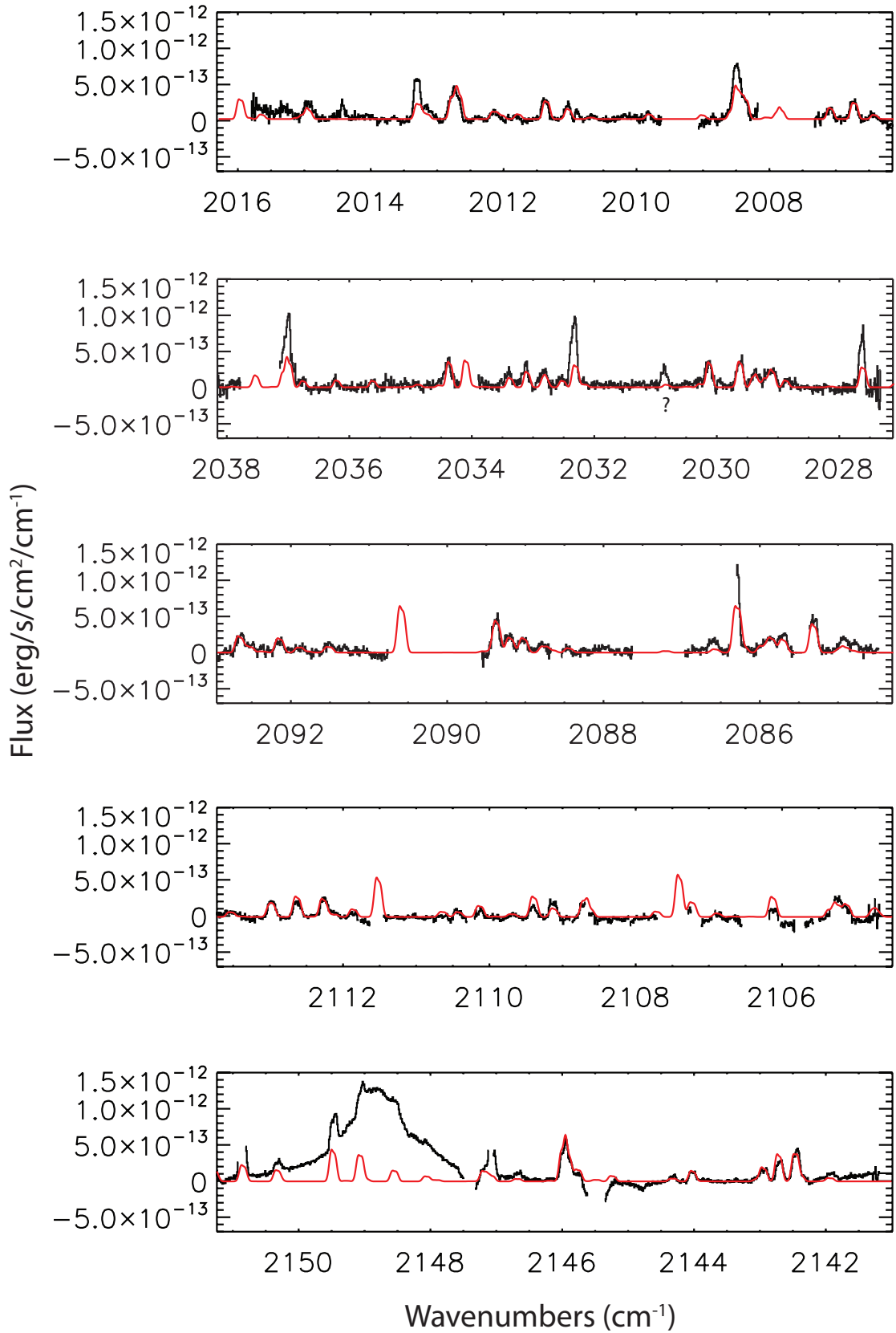}
\caption{Synthetic Spectrum of CO without collisions. The spectrum was ratioed to a telluric standard and the flux scale was determined by adopting M=3.8. The stellar continuum was subtracted from the flux calibrated spectrum before being fit with the CO model (red-solid line). The model spectrum is generated assuming the gas is only vibrationally excited by UV fluorescence. The v=1-0 high-J lines (e.g. P26 at 2032.3) are underestimated when collisional excitation is neglected. }
\end{figure}

\clearpage

\begin{figure} 
\epsscale{0.7}
\plotone{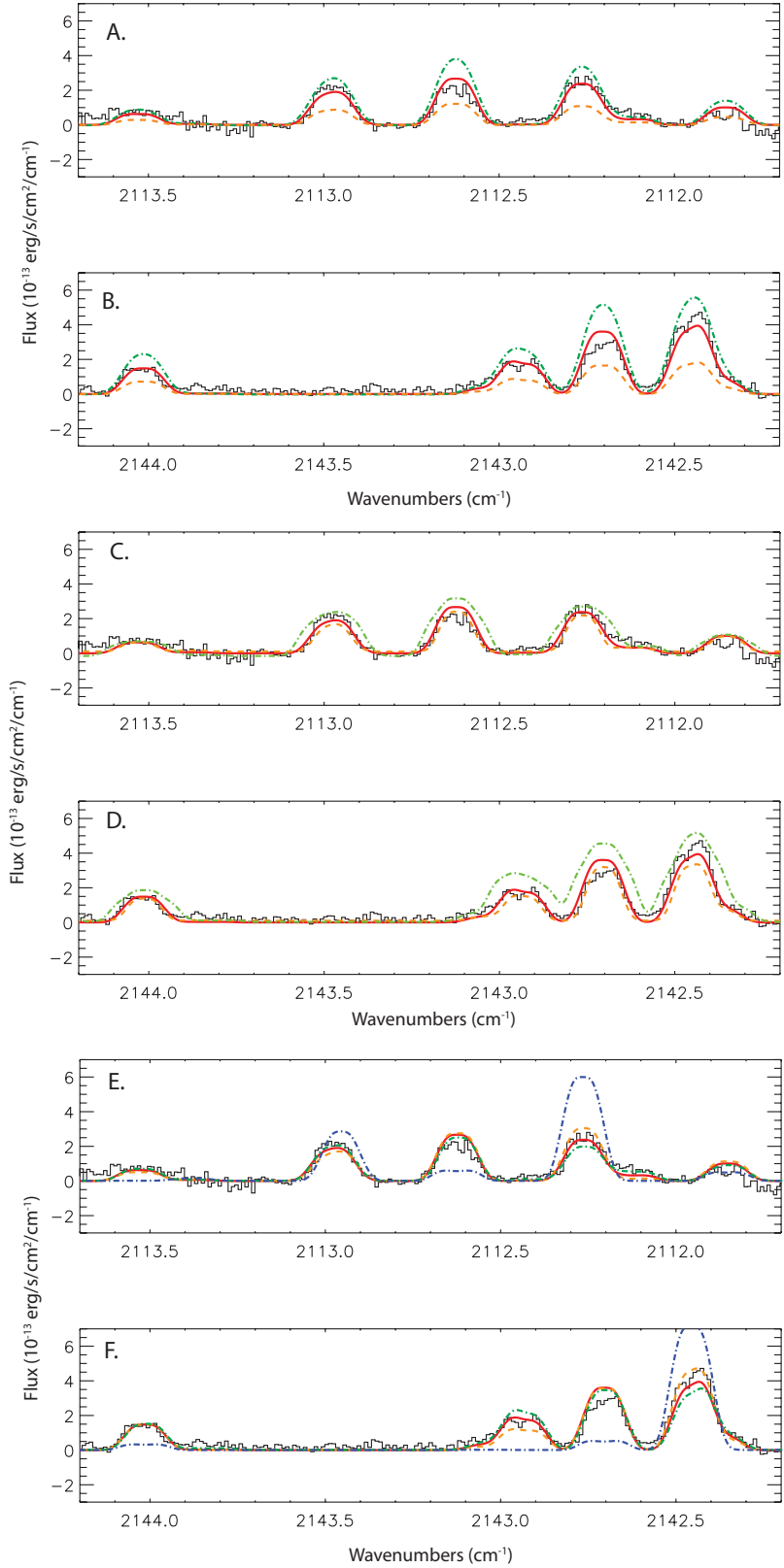}
\caption{Comparison of parameters. In panels A and B we compare the effect of the turbulent broadening parameter $b$ for values of 1 km s$^{-1}$ (orange dashed line), 2 km s$^{-1}$ (red solid line) and 3 km s$^{-1}$ (green dot-dashed line). In panels C and D we compare the effect of placing the inner edge of the disk at 7 AU (green dot-dashed line), 13 AU (red solid line) and 19 AU (orange dashed line). In panels E and F we compare the effect of the temperature of the inner disk for values of 200 K (blue dot-dashed line), 1000 K (orange dashed line), 1400K (red solid line), and 1800K green dot-dashed line).} 
\end{figure}

\clearpage

\begin{figure} 
\epsscale{0.9}
\plotone{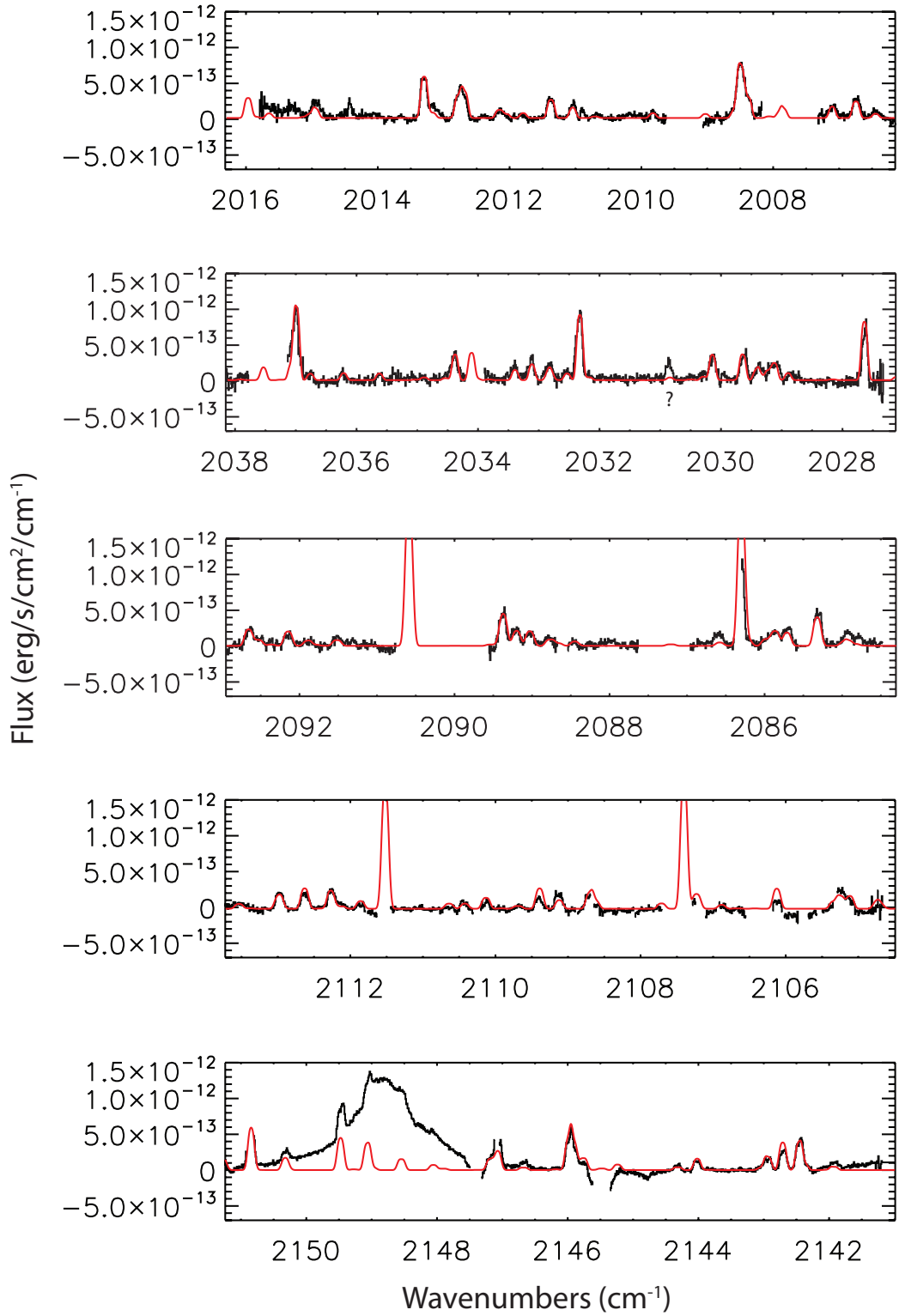}
\caption{Synthetic Spectrum of CO with collisions. Adding collisional excitation to our model improves the fit of the v=1-0 lines such as the P26 transition.} 
\end{figure}

\clearpage

\begin{figure} 
\epsscale{1.0}
\plotone{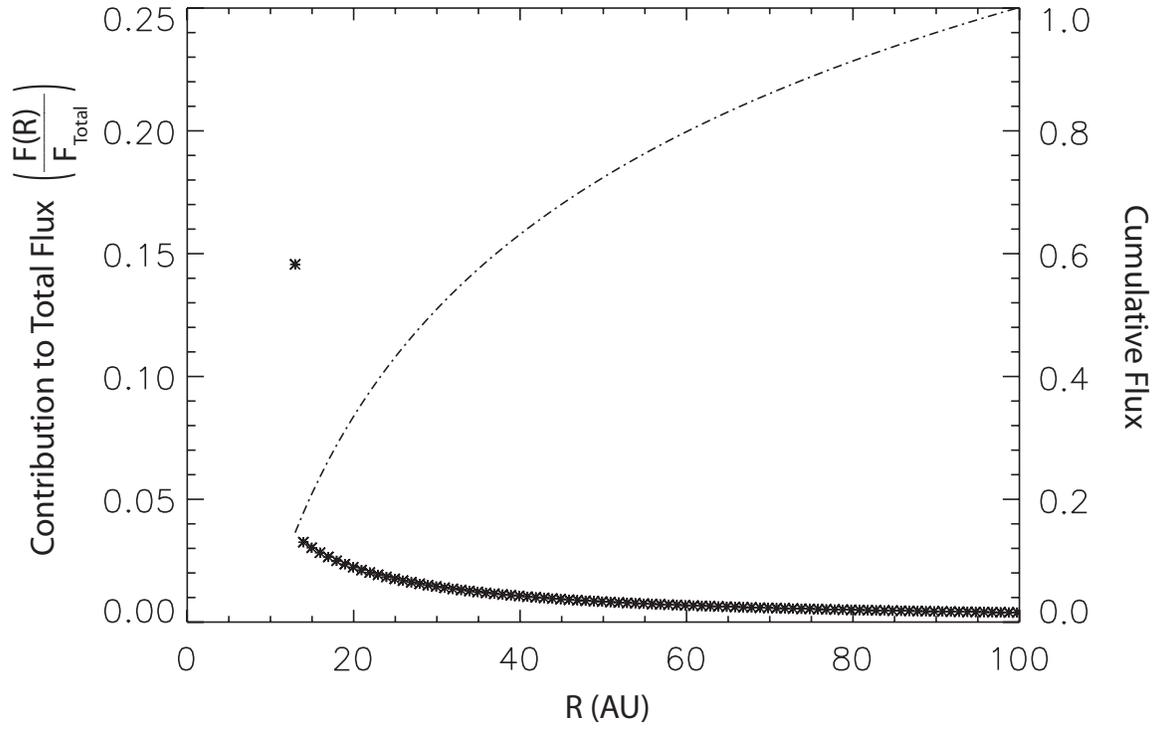}
\caption{Contribution to flux as a function of radius. Nearly half (47\%) of the flux comes from beyond 30 AU. Thus we can detect gas in disks truncated at distances comparable to the Kuiper belt in our own Solar System.} 
\end{figure}

\begin{figure} 
\epsscale{1.0}
\plotone{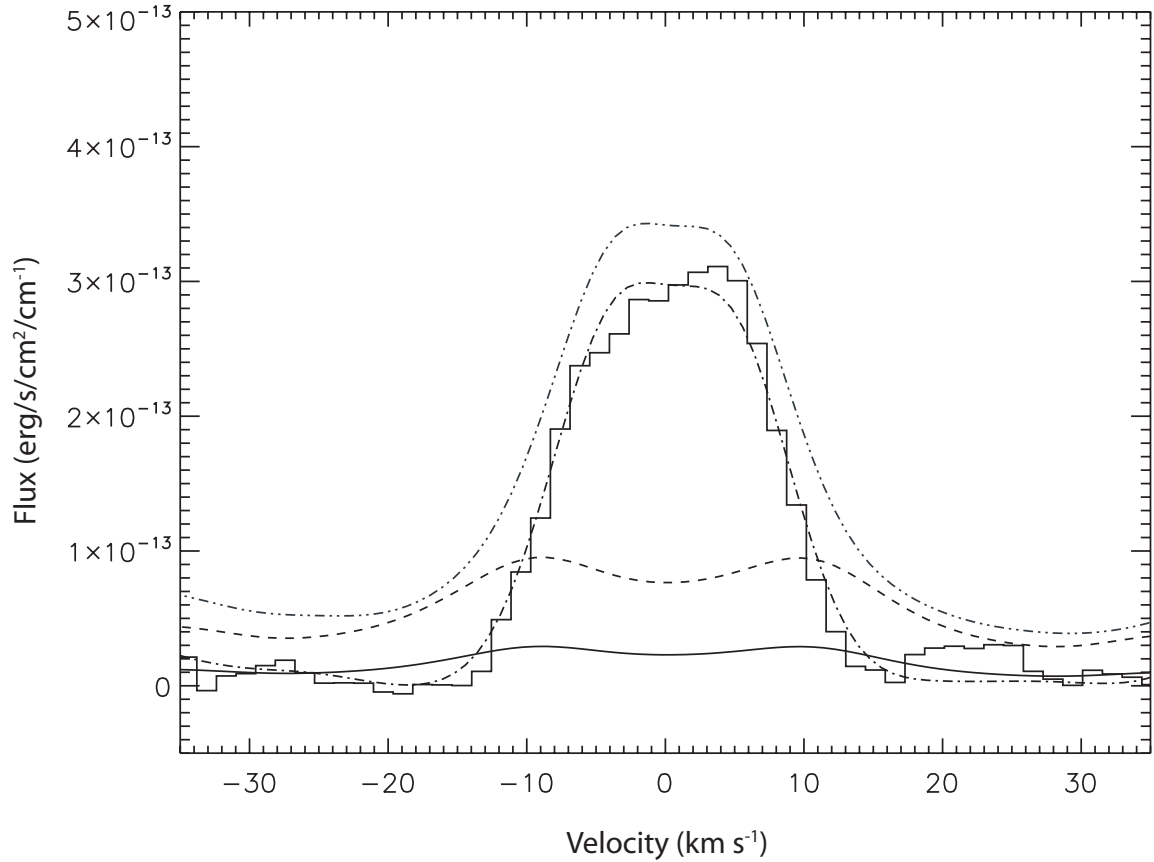}
\caption{Synthetic spectra of the average line profile. We plot the flux-calibrated average CO line profile in velocity space. The same lines presented in figures 2 and 3 were used to generate this plot. The lines were not normalized as they were for the plot presented in figure 3. Overplotted are four synthetic spectra. The dot-dashed line is the average profile of CO produced by our model using the best-fit parameters presented in table 4. The dashed line is the average profile of CO produced by gas extending from 0.3-13AU and surface density of  1.3 $\times$ 10$^{14}$ cm$^{-2}$ which we take to be our 3$\sigma$ upper limit. The solid line reflects our 1$\sigma$ upperlimit of 4.4$\times$10$^{13}$cm$^{-2}$. The dot-dot-dashed line is the spectrum of gas extending from 0.3-100 AU. }
\end{figure}

\begin{figure} 
\epsscale{0.9}
\plotone{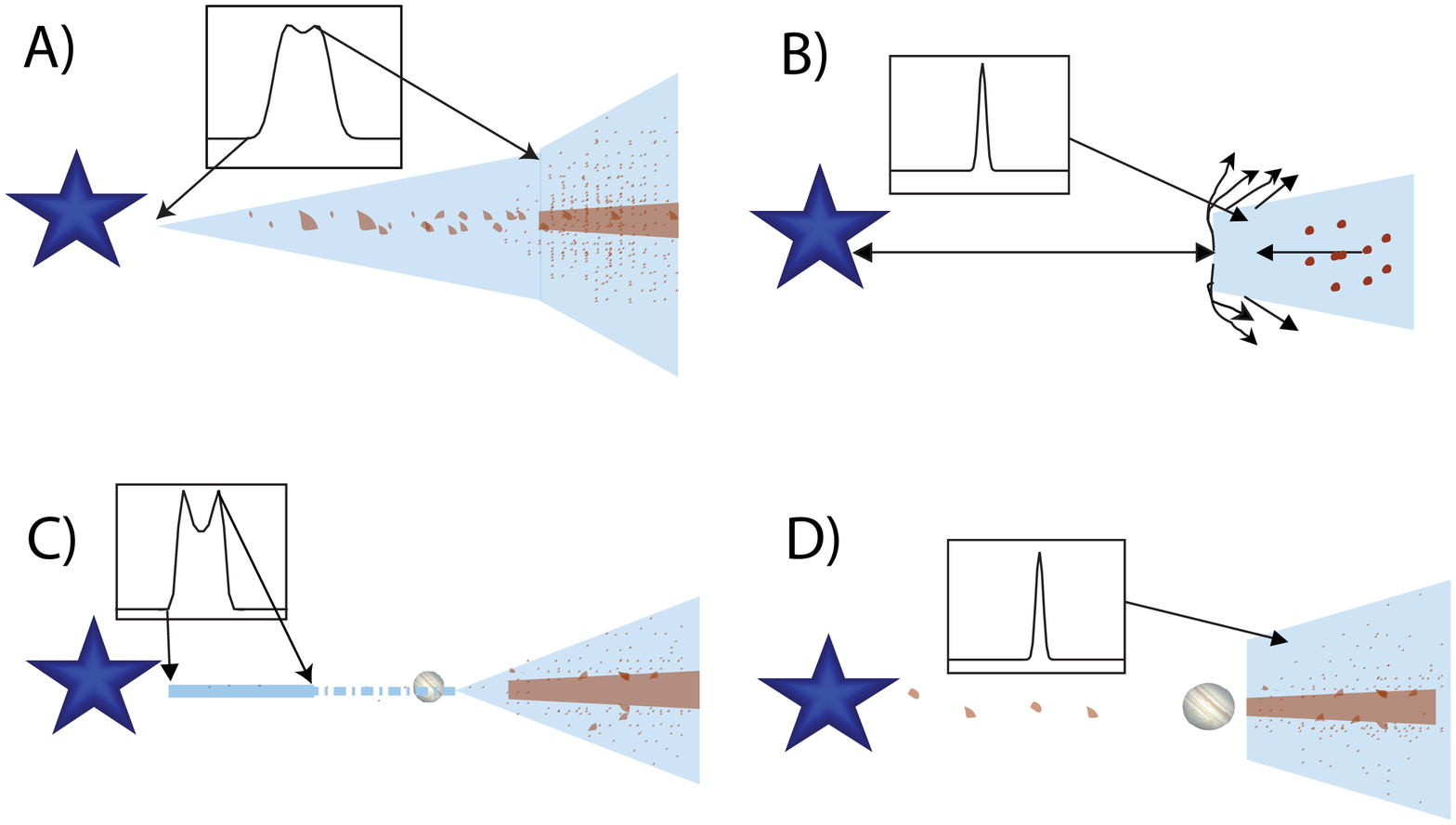}
\caption{The gas phase signature of transitional disks formed by different effects. Each panel illustrates a physically realistic scenario that can give rise to transitional SEDs. A) The preferential formation of planetesimals in the inner disk will result in gas emission from the inner disk; B) The photoevaporation of the disk will result in no gas emission from the inner disk and possibly weak emission from the outer disk; C) The presence of a jovian mass planet will result in warm gas emission over a restricted region of the inner disk; D) The presence of a suprajovian mass planet (M$\ge$5M$_{\rm Jupiter}$) will result in emission from the outer disk. B and D can be distinguished by the surface density of the outer disk inferred by millimeter observations.} 
\end{figure}

\end{document}